\documentclass[letterpaper, 10 pt, conference]{ieeeconf}
\pdfoutput=1

\usepackage[backend=bibtex, style=numeric]{biblatex}
\bibliography{./bibliography}

\newcommand\restr[2]{{
  \left.\kern-\nulldelimiterspace 
  #1 
  \vphantom{\big|} 
  \right|_{#2} 
  }}

\usepackage{amsmath, bm}
\usepackage[braket,qm]{qcircuit}
\newcommand{\gene}[1]{\gamma_{#1}}
\newcommand{\geneinstance}[2]{\gene{#1}^{#2}}
\newcommand{\genome}[1]{\bm{g}_{#1}}

\usepackage{graphics} 

\title{\bf
A Domain-agnostic, Noise-resistant, Hardware-efficient\\ Evolutionary Variational Quantum Eigensolver
}
\usepackage[mathlines, switch]{lineno}
\usepackage{booktabs}
\usepackage{amsmath}
\usepackage{graphicx}
\usepackage{fixltx2e}
\usepackage{amssymb}
\usepackage{amsfonts}
\usepackage{float}
\usepackage{afterpage}
\usepackage{subfig} 
\usepackage{caption}
\usepackage{multicol}

\newcommand{\gateset}{\mathcal{G}}
\newcommand{\optnoparams}{\mathcal{O}}
\newcommand{\opt}[1]{\optnoparams(#1)}
\usepackage{tikz}
\usetikzlibrary{shapes.gates.logic.US,trees,positioning,arrows}

\usepackage{mathtools}
\DeclarePairedDelimiter{\ceil}{\lceil}{\rceil}

\usepackage{stmaryrd}
\newcommand{\indicator}[1]{\llbracket #1 \rrbracket}
\newcommand{\angstrom}{\textup{\AA}}
\begin{document}

\author{
Arthur G. Rattew,$^{1,3}$ Shaohan Hu,$^{1}$ Marco Pistoia,$^{2}$ Richard Chen$^{1}$ and Steve Wood$^{1}$\\
\textit{$^1$IBM T. J. Watson Research Center, Yorktown Heights, NY USA}\\
\textit{$^2$JPMorgan Chase \& Co.}\\
\textit{$^3$Washington University in St. Louis}
\vspace{-3mm}
}

\maketitle
\thispagestyle{empty}
\pagestyle{empty}

\begin{abstract}
Variational quantum algorithms have shown promise in numerous fields due to their versatility in solving problems of scientific and commercial interest. However, leading algorithms for Hamiltonian simulation, such as the Variational Quantum Eigensolver (VQE), use fixed preconstructed ansatzes, limiting their general applicability and accuracy. Thus, variational forms---the quantum circuits that implement ansatzes ---are either crafted heuristically or by encoding domain-specific knowledge. In this paper, we present an Evolutionary Variational Quantum Eigensolver (EVQE), a novel variational algorithm that uses evolutionary programming techniques to minimize the expectation value of a given Hamiltonian by dynamically generating and optimizing an ansatz. The algorithm is equally applicable to optimization problems in all domains, obtaining accurate energy evaluations with hardware-efficient ansatzes.  In molecular simulations, the variational forms generated by EVQE are up to $18.6\times$ shallower and use up to $12\times$ fewer CX gates than those obtained by VQE with a unitary coupled cluster ansatz. EVQE demonstrates significant noise-resistance properties, obtaining results in noisy simulation with at least $3.6\times$ less error than VQE using \textit{any} tested ansatz configuration. We successfully evaluated EVQE on a real 5-qubit IBMQ quantum computer. The experimental results, which we obtained both via simulation and on real quantum hardware, demonstrate the effectiveness of EVQE for general-purpose optimization on the quantum computers of the present and near future.
\end{abstract}

\section{Introduction}
\label{section:introduction}
Current quantum hardware belongs to the class of Noisy Intermediate-scale Quantum (NISQ) computers~ \cite{Preskill2018}. These devices are primarily limited by 2-qubit-gate fidelity and qubit-coherence times. Consequently, many quantum algorithms necessitating deep circuits with many 2-qubit operations are not feasible for execution on the hardware of the foreseeable future. Motivated by finding commercial applications for NISQ hardware, hybrid quantum/classical algorithms, which potentially offer solutions to classically intractable problems, are being widely explored. Such quantum algorithms could have significant commercial applications in numerous fields, including quantum chemistry, logistics, healthcare and finance~\cite{Peruzzo2014, Farhi2014, Mohseni2017, Dash2019}. Moreover, by creating economic demand for extant quantum devices, such algorithms could enable sustained corporate investment in quantum-computing technology, creating a virtuous cycle resulting in rapid progress mirroring that of the semiconductor industry~\cite{National2019VirtuousCycle}.

In 2004, Peruzzo \textit{et al.} proposed a quantum/classical hybrid algorithm called the Variational Quantum Eigensolver (VQE), which, compared to previous quantum algorithms, substantially reduced circuit depth at the cost of increasing the required number of circuit executions~\cite{Peruzzo2014}. Through application of the variational method of quantum mechanics, the algorithm bounds the ground-state energy, $E_{gs}$, of a system described by a Hamiltonian $H$. The variational principle states that the expectation value of $H$ over an arbitrary normalized wave function $\ket{\psi}$ cannot be lower than the system's ground-state energy, or, more formally:
\[
    E_{gs} \le \bra{\psi}H \ket{\psi}.
\]

VQE uses a fixed circuit containing parameterized gates, whose parameters are represented with $\vec{\theta}$, to generate an ansatz $|\psi(\vec{\theta})\rangle$, over which the expectation value of $H$ is taken. Through iterative classical optimization of the parameters, the algorithm bounds $E_{gs}$ as follows:
\[
    E_{gs} \le \min_{\vec{\theta}}\left(\langle\psi(\vec{\theta})| H |\psi(\vec{\theta})\rangle \right)
\]
with the hope that the resulting expectation value closely approximates the ground-state energy of the Hamiltonian. Moreover, the applicability of VQE can be generalized through the creation of an Ising Hamiltonian to represent a wide range of optimization problems. 

Unfortunately, the accuracy of solutions generated by VQE are limited by its use of fixed variational forms. The number of degrees of freedom in a quantum system is exponential in the number of qubits. Thus, to generate a mapping to any state in the Hilbert space, a fixed parameterized circuit must have a number of parameters that is exponential to the number of qubits. However, to allow tractable classical optimization, the number of parameters in variational forms are kept polynomial in the number of qubits. Therefore, VQE can only generate transformations to an exponentially small subspace of an $n$-qubit Hilbert space.

To circumvent this limitation, VQE is often used in tandem with specially crafted domain-specific variational forms, utilizing prior knowledge about the target Hamiltonian's possible ground states.  This is, for example, the case of the Unitary Coupled Cluster for Single and Double excitations (UCCSD) variational form, commonly used in quantum-chemistry computations~\cite{Barkoutsos2018}. The creation of such variational forms is challenging, and often does not produce optimal circuits for the task in terms of accuracy, depth, and number of 2-qubit gates. Furthermore, these circuits are often crafted independently of the hardware on which they are to be executed. Together, these limitations severely hinder the practicality of VQE as a general-purpose, quantum-enabled, optimization system in the NISQ era.

In this paper, we propose the Evolutionary Variational Quantum Eigensolver (EVQE) algorithm, which addresses the same problems as VQE while compensating for some of its well-known drawbacks. We do so by utilizing evolutionary programming techniques to adaptively and concurrently search the space of circuit forms and their parameterizations. This enables EVQE to develop efficient structures automatically customized to the given problem instance. In summary, EVQE has the following novel characteristics:

\begin{enumerate}
    \item EVQE operates in a domain-agnostic fashion, making it equally applicable to problems in diverse fields, such as chemistry, optimization, finance and artificial intelligence, thereby alleviating the need for domain-specific variational forms, whose construction requires specialized knowledge.
    \item The circuits automatically generated by EVQE are significantly shallower, and use substantially fewer 2-qubit gates, while achieving comparable if not better results than those produced by alternative domain-specific algorithms.
    \item EVQE is quantum-hardware adaptive; it automatically favors the construction of circuits that are more resilient to the noise characteristics and connectivity constraints of the specific quantum computer on which the algorithm is executed.
\end{enumerate}
More specifically, EVQE ranges from having $5.0\times$ to $15.2\times$ shallower circuits, using $3.5\times$ to $5.0\times$ fewer CX gates, than the VQE/UCCSD configuration when estimating the ground-state energy of LiH. In the estimation of the ground-state energy of BeH$_2$, EVQE obtains $18.7\times$ shallower circuits, using $12.0\times$ fewer CX gates than VQE/UCCSD. In the maximum-cut and vehicle-routing problems, EVQE obtains efficient and optimal results more consistently than VQE. In the noisy simulation of LiH, EVQE obtains results with \textit{at least} $3.4\times$ less error than the best VQE results. Finally, EVQE is successfully demonstrated on a real 5-qubit IBMQ quantum processor.

\section{Background}
\label{section:background}

VQE employs the variational method of quantum mechanics to bound the ground-state energy of a Hamiltonian. As already discussed, VQE is limited by its selection of a variational form---a parameterized trial function---in that an efficiently parameterized fixed variational form is unable to produce a mapping to the ground state of an arbitrary Hamiltonian.

Consequently, a method to vary the form of a variational circuit is required to capture every possible mapping with an efficient number of parameters. This reduces the problem of finding the ground state of an arbitrary Hamiltonian to selecting an efficient set of 1- and 2-qubit parameterized gates that map to that ground state. Assuming that an optimal parameterization of such a circuit could be obtained, the ground state energy of an arbitrary Ising Hamiltonian could be found, and so either this problem is NP-Hard or it is not possible to efficiently optimize a fixed variational circuit. Thus, heuristic methods may be required in practice. For example, the algorithm \textit{ADAPT-VQE} utilizes knowledge from coupled-cluster theory to adaptively construct and optimize variational circuits for molecular simulations~\cite{Grimsley2019}. However, ADAPT-VQE has several limitations---such as a likely sensitivity to noise---which may hinder its applicability on NISQ devices. In 2019, Ostaszewski \textit{et al.} presented two greedy variational algorithms, one of which adaptively grows the variational circuits it uses~\cite{Ostaszewski2019}. However, that algorithm only adaptively adds single-qubit gates, leaving the 2-qubit gates fixed, and thus has similar limitations as VQE. 

A variational algorithm that efficiently grows and optimizes its parameterized forms would also have applications in machine learning. For example, there is a natural similarity between the approximation of functions in classical machine learning---such as with neural networks or kernel methods---and the variational minimization of quantum circuits. Consequently, various hybrid quantum/classical machine-learning algorithms have been proposed that utilize fixed variational forms to implement their function-approximation systems~\cite{Mitarai2018, Schuld2018, Havlivcek2019, Havlivcek2019}. The classical counterparts to each of these algorithms have fixed parameter sets, and so a correspondence to the parameters in a fixed variational circuit is natural. By contrast, in 2002, Stanley and Miikkulainen proposed an algorithm titled NeuroEvolution of Augmenting Topologies (NEAT), which employs a genetic algorithm to concurrently grow and optimize neural networks \cite{Stanley2002}. As the form of its neural networks vary, its parameterizations change, and so it is not analogous to utilizing a fixed variational form. Nevertheless, some of the techniques described by Stanley and Miikkulainen transition to the quantum setting, and mirror some of those used by EVQE.

While various genetic algorithms have been proposed to evolve a quantum circuit corresponding to a target matrix, to the best of our knowledge, none are directly applicable to variational minimization and none explicitly focus on concurrently evolving and optimizing parameterized circuits~\cite{Williams1998,Rubinstein2001,Lukac2003,Ding2008,Wang2014}. Moreover, these algorithms primarily use the crossover genetic operator to explore their respective search spaces, categorizing them as sexual genetic algorithms. Crossover fuses the genomes of two parents to produce an offspring, for example, by concatenating a portion of each parent's corresponding quantum circuit. However, because of entanglement between qubits and the superposition of quantum states, merging the circuits of two parent genomes to produce an offspring does not necessarily produce a circuit whose mapping is similar to that of either parent, even if the parents are closely related. Thus, although the resulting circuit form may be preferable, its parameters would still need to be re-optimized, wasting the optimization iterations performed on all of its ancestors. Furthermore, these algorithms primarily grow circuits using non-parameterized gates. Therefore, when non-identity gates are added to an existing circuit, because of superposition and entanglement, the energy evaluation of the overall circuit changes non-smoothly. This substantially increases the challenge of minimizing the expectation value of a Hamiltonian. In contrast, EVQE explores its solution space utilizing evolutionary programming techniques, meaning that each member of the population has only one parent, and offspring are differentiated primarily through random mutations. This enables EVQE to circumvent the issues associated with fusing two disparate circuits, and more significantly, to smoothly and efficiently explore the Hilbert space, as explained in the next section. Moreover, combined with some of EVQE's other characteristics, this identity-initialized growth mechanism likely enables EVQE to circumvent a significant problem associated with variational quantum algorithms that utilize random circuits. 

First identified by McClean \textit{et al.} in 2018, the \textit{barren-plateau problem} states that random quantum circuits with sufficient depths experience an exponentially small probability of having non-zero parameter gradient readings, in terms of the number of qubits~\cite{Mcclean2018}. Thus, there is no clear way to optimize the parameters in such variational circuits, which prevents them from efficiently scaling. However, EVQE has a number of mechanisms that help it avoid the barren-plateau problem. First, its circuits are initialized with a single random layer. As mentioned by McClean, the barren-plateau problem only appears in random circuits with much greater depths. However, should this still be not desirable, it would be easy to specify EVQE's starting population of circuits such that the starting population is outside of a barren region. Either way, subsequent circuit growth occurs with identity-initialized parameters, which allows EVQE to avoid entering regions in which the gradient is imperceptible. This is similar to the strategy proposed by Grant \textit{et al.} in 2019, which also addresses this problem~\cite{Grant2019}. 

\section{Evolutionary Variational Quantum Eigensolver (EVQE)}

EVQE is a speciated, asexual, evolutionary algorithm for general-purpose multimodal optimization. By mirroring the processes of natural selection, the algorithm effectively explores a search space of quantum-circuit forms and parameterizations. This section provides a brief overview of EVQE, and summarizes the characteristics that enable its novel properties.

\subsection{Genetic Representation of Quantum Circuits}
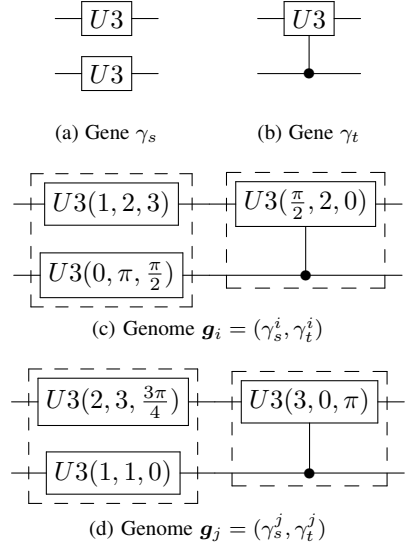
\begin{figure}[t]
    \captionsetup{font=small}
    \centering
    \subfloat[Gene $\gene{s}$]{$\Qcircuit @C=1em @R=.8em {
                & \gate{U3}           & \qw  \\
                & \gate{U3} & \qw \\
                \vspace{.1cm}
            }$} \hspace*{1.2cm}
    \subfloat[Gene $\gene{t}$]{$\Qcircuit @C=1em @R=1.32em {
                & \gate{U3}           & \qw  \\
                & \ctrl{-1} & \qw \\
                \vspace{.1cm}
            }$}

    \subfloat[Genome $\genome{i}=(\geneinstance{s}{i},\geneinstance{t}{i})$]{$\Qcircuit @C=1em @R=1em {
        & \gate{U3(1, 2, 3)}           & \qw & \gate{U3(\frac{\pi}{2}, 2, 0)}  & \qw \\
        & \gate{U3(0, \pi, \frac{\pi}{2})} & \qw & \ctrl{-1}    & \qw \gategroup{1}{2}{2}{2}{.7em}{--}\gategroup{1}{4}{2}{4}{.7em}{--}
    }$}\\

    \subfloat[Genome $\genome{j}=(\geneinstance{s}{j},\geneinstance{t}{j})$]{$\Qcircuit @C=1em @R=1em {
                & \gate{U3(2, 3, \frac{3\pi}{4})}           & \qw & \gate{U3(3, 0, \pi)}  & \qw \\
                & \gate{U3(1, 1, 0)} & \qw & \ctrl{-1}    & \qw \gategroup{1}{2}{2}{2}{.7em}{--}\gategroup{1}{4}{2}{4}{.7em}{--}
            }$}
    \caption{\textbf{Illustration of Genes, Gene Instances and Genomes}: Two possible genes, $\gene{1}$ and $\gene{2}$, as well as two genomes, $\genome{i}$ and $\genome{j}$, containing instances of those genes, are shown. Genes fully specify the gates, not the parameters, of a layer of a quantum circuit. Gene instances, the elements of the list constituting a genome, are instantiations of a gene which specify all applicable parameters.}
    \label{fig:genetic_illustration}
\end{figure}

At a high level, the algorithm maintains a population of genomes that represent quantum circuits. A \textit{genome} $\genome{i}$ is a list of \textit{genes}, where each gene fully describes  a layer of a quantum circuit. A gene is represented by $\gene{\alpha}$, where $\alpha$ is a unique identifier of that gene. A \textit{gene instance}, $\geneinstance{\alpha}{i}$, describes an instance of $\gene{\alpha}$ found in a genome, $\genome{i}$. For example, a genome $\genome{i}$ with $m$ genes may be represented as $\genome{i} = (\geneinstance{\alpha_1}{i}, \geneinstance{\alpha_2}{i}, ..., \geneinstance{\alpha_m}{i})$. This is illustrated in Figure \ref{fig:genetic_illustration}.

A gene, $\gene{\alpha}$, characterizes a layer of a quantum circuit, such that each qubit in that layer is assigned a gate from the following set:
\[
    \gateset = \{\mathbb{I}_2, \textup{U3}, \land_1(\textup{U3}) \}.
\]
Here, $\mathbb{I}_2$ is the identity gate, U3 is a universal
single-qubit gate with 3 parameters, and $\land_1(\textup{U3})$ represents a controlled-U3 gate, often indicated as CU3 as well~\cite{Barenco1995}. Furthermore, a gene instance contains all of the parameters required for any parameterized gates it describes. 

\subsection{Asexual Reproduction and Speciation Through Genetic Ancestry}

Since EVQE uses an asexual reproduction scheme, its primary method of exploring the solution space is mutation, in contrast to typical sexual genetic algorithms where crossover is the preferred exploration operation, as mentioned in the background section. The asexual reproduction scheme enables EVQE to optimize clearly-defined gradients, efficiently identify niches for speciation, and circumvent the permutation problem. Asexual reproduction also contributes to the algorithm's noise resistance. 

As shown by Goldberg in 1989, and demonstrated in the domain of neural networks by Angeline, \textit{et al.} in 1994, crossover is most effective in genetic algorithms when the performance of a genome is related to that of its constituent components~\cite{Angeline1994, Goldberg1989}. However, because of the entanglement and superposition of qubits, the cost evaluation of a subset of a quantum circuit is not clearly related to the cost evaluation of the overall circuit. We are going to elaborate on this next.

Unlike previous work using genetic algorithms to grow quantum circuits (given a known target matrix), when EVQE adds a gate to a quantum circuit, that gate's parameters are initialized such that the gate performs the identity transformation~\cite{Williams1998,Rubinstein2001,Lukac2003,Ding2008,Wang2014}. Thus, when a new gate is added to a circuit, that circuit's expectation value is unchanged. The parameters of new gates are altered through optimization, so the addition can only reduce the circuit's expectation value. In contrast, a system using crossover between two genomes would be unable to preserve the expectation value of either circuit, hindering the algorithm's ability to consistently improve, and thus converge, and would require all of the parameters in the new circuit to be re-optimized, thereby wasting all previous optimization progress.

Moreover, asexual reproduction enables an effective speciation scheme. \textit{Speciation}, also known as \textit{niching}, enables genetic algorithms to maintain a diverse set of candidate solutions in the population. This greatly improves the ability of those algorithms to optimize multi-modal functions as the species in the population concentrate in various optima throughout the search space~\cite{Mahfoud1995}.

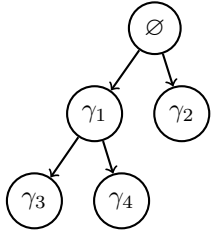
\begin{figure}[h]
    \captionsetup{font=small}
    \centering
    \begin{tikzpicture}[scale=0.1]
        \begin{scope}[thick, minimum size=0.5cm, node distance=0.4cm,  comp/.style={circle, draw, fill=white}]
           \node [comp]	 (cb)                   {$\varnothing$};
           \node [comp]	 (ca1)	[below=of cb,xshift=-0.8cm]	{$\gene{1}$} edge [<-] (cb);
           \node [comp]	 (ca2)	[right=of ca1]			{$\gene{2}$} edge [<-] (cb);
           \node [comp]	 (ca11)	[below=of ca1,xshift=-0.8cm]	{$\gene{3}$} edge [<-] (ca1);
           \node [comp]	 (ca12)	[right=of ca11]			{$\gene{4}$} edge [<-] (ca1);
       \end{scope}
    \end{tikzpicture}
    \caption{\textbf{Genetic Representation of a Population of Quantum Circuits.} A genetic tree corresponding to a population with 3 genomes is shown. Each of the genomes in the population contains part of the information required to construct this tree. Moreover, this tree illustrates the set of circuit forms that have been explored by the ancestors of the current population.}
    \label{fig:genetic_tree}
\end{figure}

Speciation requires a metric to determine the \textit{genetic distance}, or \textit{similarity}, of any two genomes, thereby enabling similar genomes to be grouped into the same species. Genetic distance should be defined such that any genes competing in the same niche are in the same species. EVQE takes inspiration from the historical innovation system proposed by Stanley and Miikkulainen in 2002, where each new gene is assigned a globally unique identifier~\cite{Stanley2002}. Whereas Stanley and Miikkulainen essentially calculate the genetic distance between two genomes as the ratio between matching and non-matching genes, EVQE uses a calculation enabled by its asexual reproduction system. In EVQE, every genome has exactly one parent. Consequently, using only the genomes in the population, an abstract genetic tree may be constructed that represents each genome's genetic ancestry. For example, the tree shown in Figure~\ref{fig:genetic_tree} represents a population with 3 genomes, $\genome{1} = (\geneinstance{1}{1}, \geneinstance{3}{1})$, $\genome{2} = (\geneinstance{1}{2}, \geneinstance{4}{2})$, and $\genome{3} = (\geneinstance{2}{3})$.

Thus, we define the genetic distance $\delta_{ij}$ between $\genome{i}$ and $\genome{j}$ to be the total number of genes in both genomes, subtracted by the number of genes shared in common. This gives the number of genes to their most recent common ancestor, and may be calculated as follows:
\[
    \delta_{ij} = \ceil[\Big]{\frac{1}{2}(|\genome{i}| + |\genome{j}|)} - \left(\sum_{k = 1}^{|\genome{i}|} \indicator{\genome{i}[k] = \genome{j}[k]} \right).
\]
Here, $|\genome{i}|$ gives the number of non-null genes in $\genome{i}$ and $\genome{i}[k]$ gives the $k^{\textup{th}}$ gene in $\genome{i}$, while $\indicator{\genome{i}[k] = \genome{j}[k]}$ evaluates to $1$ if the $k^{\textup{th}}$ gene in both genomes are the same, or $0$ otherwise. 

Moreover, as detailed later, the parents for the genomes in each new generation are selected (with replacement) from the existing population, with probabilities proportional to their fitness. In expectation, asexual reproduction ensures that fit genomes in any generation have multiple offspring in the subsequent generation. As all of the immediate offspring of a genome are usually similar to each other (and produce similar energy evaluations), for noise to extinguish a promising genetic line, it must non-trivially harm the fitness evaluations of all of the genomes in that line simultaneously. That is, noise must wipe out all of the members of a genetic line concurrently or the speciation mechanism will repopulate that genetic line in subsequent generations. Assuming a simple stochastic additive noise model, the probability of this happening is exponentially small in terms of the number of members in the genetic line.

\subsection{Fitness Evaluation}
The fitness $f_i$ of genome $\genome{i}$ is a measure of how well $\genome{i}$ optimizes the objective function. We define $f_i$ to include both the energy evaluation of $\genome{i}$'s corresponding circuit, $\ket{\psi_i}$, and two penalty terms. The depth of $\genome{i}$'s circuit is simply the number of genes it contains (since each gene corresponds to a single circuit layer) and is therefore given by $|\genome{i}|$. Furthermore, allow $\textup{CU3}(\genome{i})$ to represent the number of $\land_1(U3)$ gates in the circuit $\ket{\psi_i}$. Then, the fitness of $\genome{i}$ is defined as follows:
\[
    f_i \equiv \bra{\psi_i}H\ket{\psi_i} + \alpha \cdot |\genome{i}| + \beta \cdot \textup{CU3}(\genome{i})
\]
where $\alpha, \beta \in \mathbb{R}$ and are non-negative, user-specified parameters. The coefficients $\alpha$ and $\beta$ should be set to small values such that the magnitude of $\alpha \cdot |\genome{i}| + \beta \cdot \textup{CU3}(\genome{i})$ corresponds to the desired precision of a solution. The net effect of the penalty terms is to encourage the population to develop the shallowest circuits with the fewest $\land_1(\textup{U3})$ gates (and, transitively, CX gates) possible, which still optimizes the expectation value $\bra{\psi_i}H\ket{\psi_i}$ to the desired level of precision. The fitness penalties have two primary benefits: encouraging the exploration of resource-efficient circuits, and mitigating the effects of noise in circuit evaluation. It is clear how the penalties enable the first benefit. Noise resistance is enabled by the penalties working similarly to regularization in machine-learning algorithms that learn from data. By adding some additional penalty, those algorithms have reduced abilities to overfit noise in their training data, leading to better generalization. Similarly, by penalizing circuit depth and CX count, EVQE implicitly penalizes circuits that are more susceptible to the noise characteristics of NISQ hardware. Thus, EVQE reduces the likelihood of its genomes tending towards circuits that experience increased noise and potentially yield deceptively low energy evaluations. However, by setting $\alpha$ and $\beta$ to small values, EVQE also does not prevent circuits from being developed that naturally lower the calculated expectation value.

EVQE utilizes \textit{explicit fitness sharing}, popularized by Goldberg and Richardson in 1987, where each genome in the population shares its fitness with the other genomes in its species~\cite{Goldberg1987}. Thus, no species dominates the population, and multiple species are concurrently maintained that explore various optima in the search space. The specific mechanism used by EVQE was described by Spears in 1995 and is compatible with our formulation of genetic distance~\cite{Spears1995}. For each genome, $\genome{i}$, the fitness score is modified to yield an adjusted fitness score defined as follows:
\[
    f_i^a \equiv \frac{f_i}{|S_a|}
\]
where $|S_a|$ is the size of the species of which $\genome{i}$ is a member.

\subsection{Mutation Operators}
The mutation operators used by EVQE are \textit{topological search}, \textit{parameter search}, and \textit{removal}.

The topological-search operator creates a new gene, corresponding to a random assignment of gates from $\gateset$ to each qubit in a circuit layer subject to the pruning optimizations detailed in the Appendix, subsequently appending a corresponding gene instance to the target genome.
For a genome $\genome{i}$, the formal definition of the topological-search operator is as follows:
\begin{linenomath}
\begin{align*}
    \tau : \genome{i} &\mapsto \genome{i'}\\
    \Leftrightarrow 
    \tau : (\geneinstance{\alpha_1}{i}, ..., \geneinstance{\alpha_m}{i}) &\mapsto (\geneinstance{\alpha_1}{i}, ..., \geneinstance{\alpha_m}{i}, \geneinstance{\alpha_{m+1}}{i}).
\end{align*}
\end{linenomath}
When each gene instance is first added to a genome, its parameters are all initialized to 0 such that the action of any U3 and $\land_1(\textup{U3})$ gates are that of the identity gate. As shown in the Appendix, this optimization yields significant reductions in the number of generations required for the algorithm to converge to a solution with fixed precision.

The parameter-search operator optimizes each layer of the genome's circuit, one at a time, according to a randomly-selected ordering. Whereas the topological-search operator explores the space of circuit forms, the parameter-search operator explores the space of parameterizations for a fixed circuit form. The parameter-search operator is formally defined as follows:
\begin{linenomath}
\begin{align*}
    \pi : \genome{i} &\mapsto \opt{\genome{i}}\\
    \Leftrightarrow 
    \pi : (\geneinstance{\alpha_1}{i}, ..., \geneinstance{\alpha_m}{i}) &\mapsto \opt{\geneinstance{\alpha_1}{i}, ..., \geneinstance{\alpha_m}{i}}
\end{align*}
\end{linenomath}
where $\opt{\genome{i}}$ optimizes each $\geneinstance{\alpha_j}{i} \in \genome{i}$ in a random order. Throughout this paper, we refer to an \textit{iteration optimization count}; this refers to the maximum number of iterations the optimization subroutine may perform each time it is called on a circuit layer.

The removal operator eliminates a random number of contiguous gene instances, starting from the last gene instance in the genome. Formally, it performs the following mutation, which allows EVQE to repopulate shallower search spaces:
\begin{linenomath}
\begin{align*}
        \rho :  \genome{i} &\mapsto \genome{i'}\\
        \Leftrightarrow  \rho : (\geneinstance{\alpha_1}{i}, ..., \geneinstance{\alpha_m}{i}) &\mapsto  (\geneinstance{\alpha_1}{i}, ..., \geneinstance{\alpha_{p}}{i})
\end{align*}
\end{linenomath}
where $p \in \{1, \ldots, m\}$ is selected uniformly at random.

\subsection{Algorithm Summary}
The following is a basic outline of EVQE:
\begin{enumerate}
    \item Create a population of $P$ genomes, and apply $\tau(\genome{i})$ to each genome once. 
    \item Randomly select one genome, $\genome{i}$, from each species and add it to the \textit{species representative set}, $\mathcal{S}$. Note that the cardinality of $\mathcal{S}$ corresponds to the number of species active in the population.
    \item Run the optimization subroutine on the last gene $\geneinstance{\alpha_m}{i}$ of each $\genome{i}$, $\opt{\geneinstance{\alpha_m}{i}}$.
    \item Calculate (a) the fitness $f_i$ for each $\genome{i}$ and (b) the species-adjusted fitness $f^a_i$.
    \item Randomly select $P$ parent genomes from the population (with replacement) according to probabilities proportional to their adjusted fitness.
    \item Mutate each parent genome, by applying each mutation operator with some probability, to create the next generation.
    \item Assign each $\genome{i}$ to a specie. Specifically, this is done by mapping $\genome{i}$ to the first specie for which $\delta_{ij}$ is less than a predefined genetic distance threshold, for each $\genome{j} \in \mathcal{S}$. If no such $\genome{j}$ exists, add $\genome{i}$ to $\mathcal{S}$, thereby defining a new specie.
    \item If $G$ generations have passed, where $G$ is a given budget, return the expectation value associated with the fittest genome encountered so far. Otherwise, return to Step 2.
    
\end{enumerate}

\section{Experimental Evaluation}
All circuit depths and CX counts reported herein are obtained after optimizing each circuit with the Qiskit Terra transpiler, configured to optimization level 3. All VQE circuits have barriers removed. Moreover, in figures where error bars are shown, the error bars represent one standard deviation from the mean.
\begin{figure}[h]
    \captionsetup{font=small}
    \centering
    \subfloat{{\includegraphics[width=0.46\linewidth]{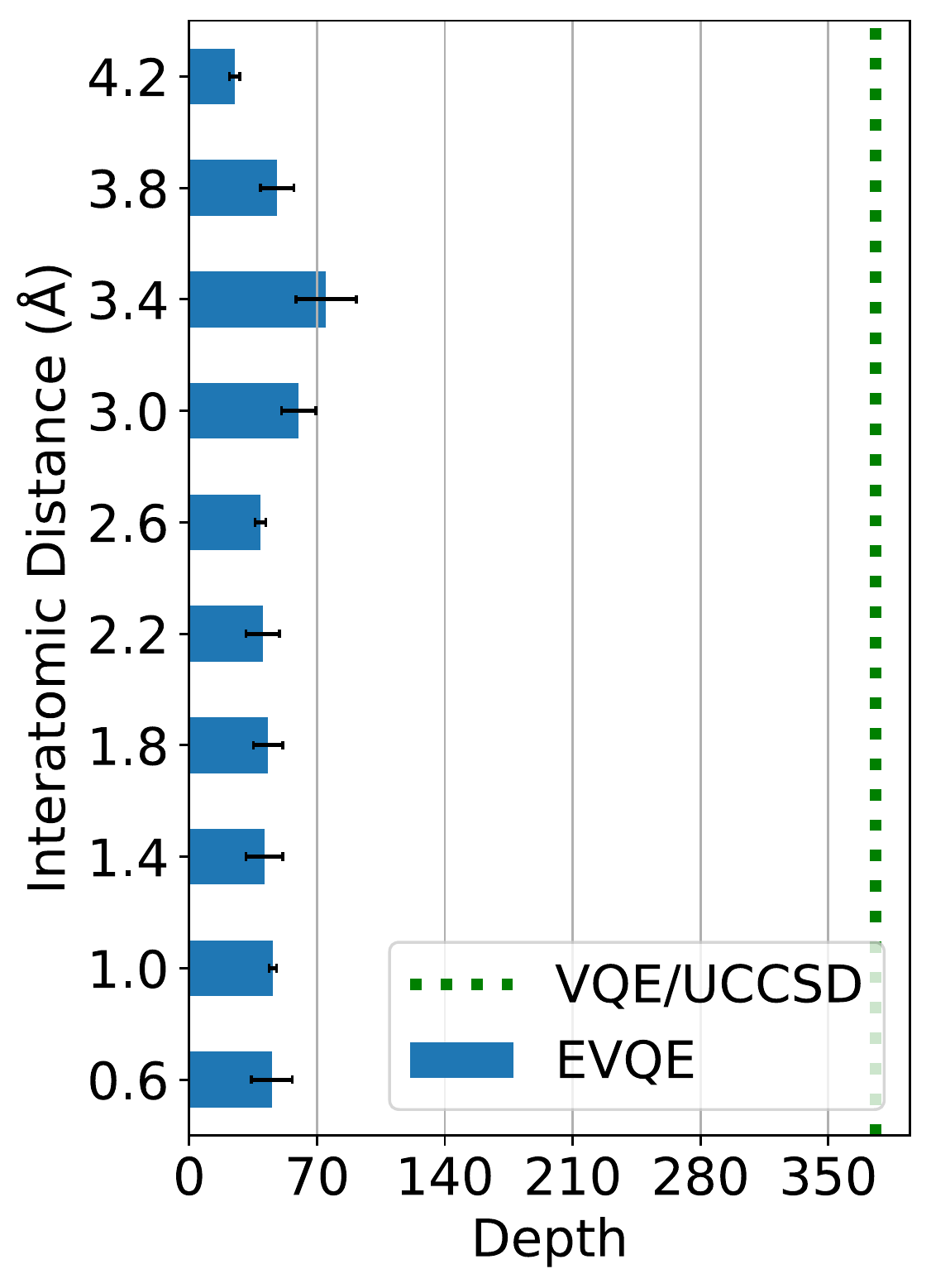}}}\hfill
    \subfloat{{\includegraphics[width=0.46\linewidth]{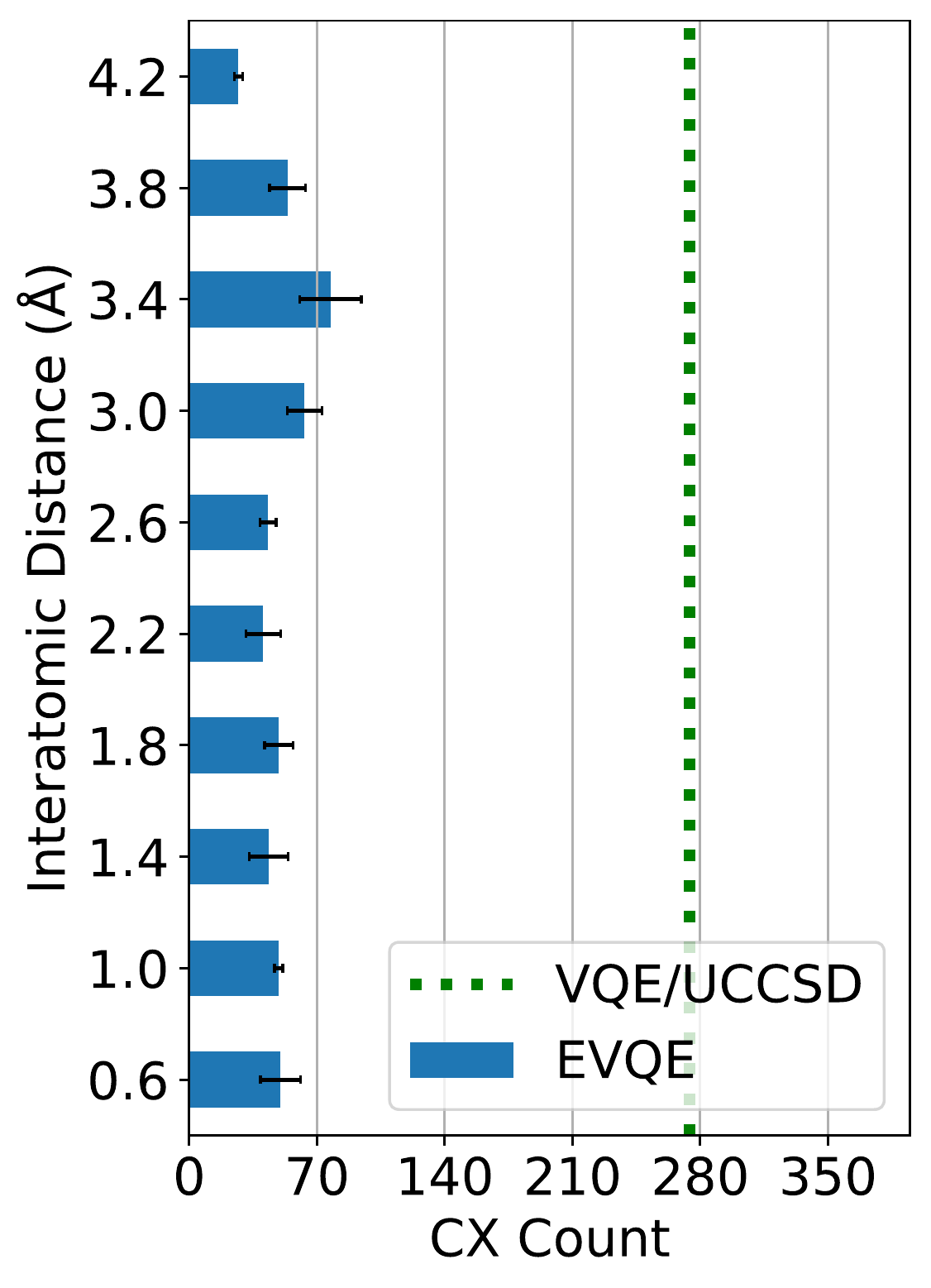}}}
    \vspace{1mm}
    \caption{\textbf{Chemistry Application: State-vector Implementation for 6-qubit LiH.} EVQE is compared against VQE/UCCSD in the estimation of the ground-state energy of LiH at various interatomic distances. EVQE is configured with a population size of 150, an optimization count of 200, an $\alpha$ of $5\times 10^{-5}$, a $\beta$ of $10^{-5}$, and the Constrained Optimization BY Linear Approximation (COBYLA) optimizer. VQE/UCCSD demonstrates the best performance when using the SLSQP optimizer. It is allowed to perform an unbounded number of optimization iterations, terminating only upon convergence. The depth setting used for VQE/UCCSD is the minimum possible. On the average of 5 trials, all algorithms obtain chemical accuracy at all interatomic distances.}
    \label{fig:lih}
\end{figure}

\label{section:experimental_evaluation}

\subsection{State-vector Simulation}

    The purpose of the experiments conducted on the state-vector simulator is to compare the performance and convergence properties of EVQE and VQE under theoretically-optimal conditions. State-vector simulation performs the matrix multiplication corresponding to a given quantum circuit, yielding the resulting quantum-state-vector. The expectation value of the matrix representing the target Hamiltonian is then taken with respect to the output state-vector.  Therefore, state-vector simulation is equivalent to noiseless simulation with infinite shots. 

\begin{figure}[t]
\captionsetup{font=small}
    \centering
    \includegraphics[width=0.84\linewidth]{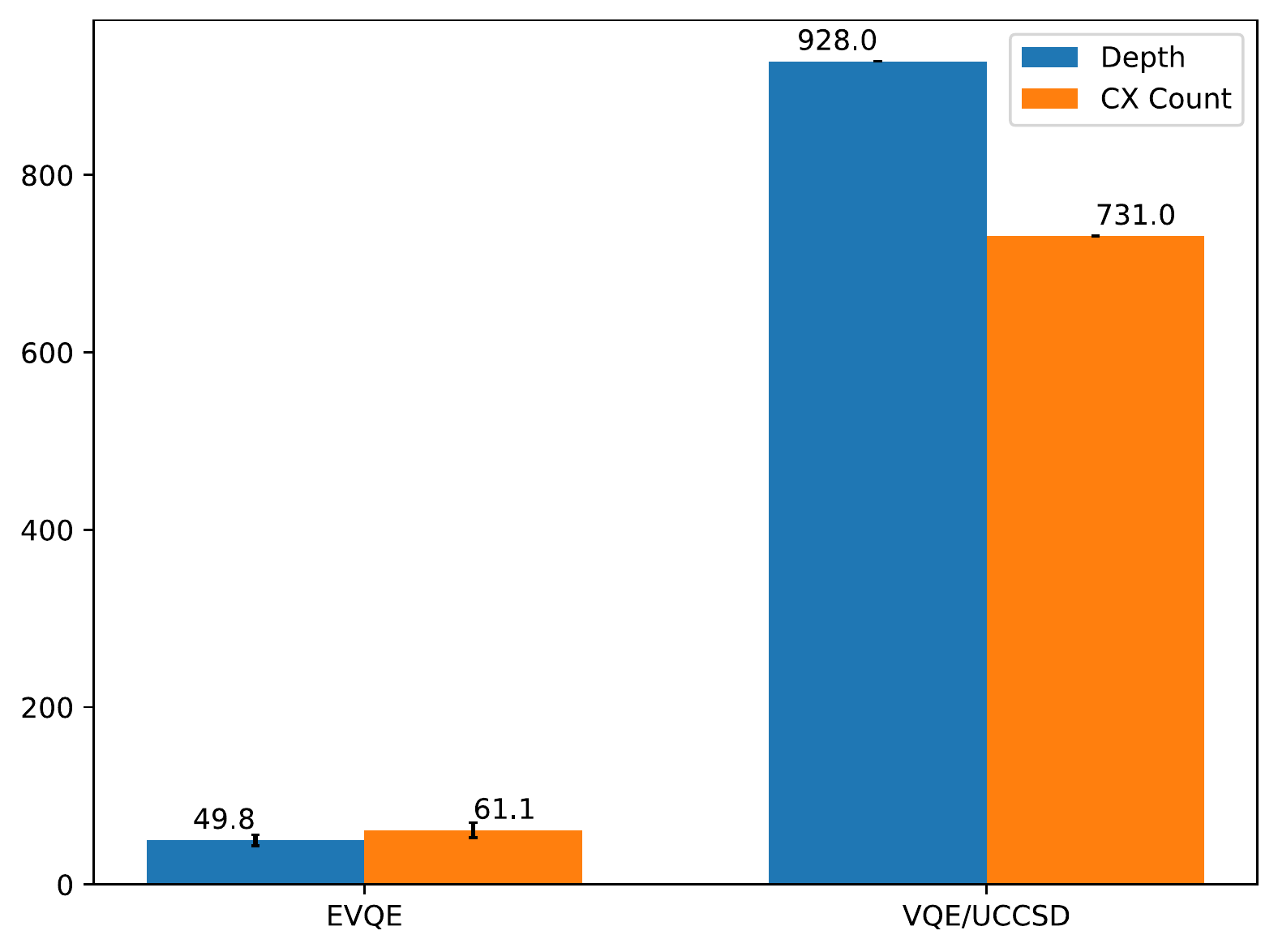}
    \caption{\textbf{Chemistry Application: State-vector Implementation for 7-qubit BeH$_2$.} EVQE is compared against VQE in the estimation of the ground-state energy of BeH$_2$ with an interatomic distance of 1.3$\angstrom$. EVQE is configured with a population size of 300, an optimization count of 300, an $\alpha$ of $10^{-5}$, a $\beta$ of $10^{-5}$, and the COBYLA optimizer. VQE demonstrates the best performance when using the SLSQP optimizer with an unbounded number of optimization iterations, terminating only upon convergence. The depth setting used for VQE/UCCSD is the minimum possible. Both algorithms obtain results within chemical accuracy over an average of 10 trials.}
    \label{fig:beh2}
\end{figure}

    The experiments conducted for this study fall into two categories: molecular simulation and general optimization.

    In the two molecular simulations presented, the task is to find the ground-state energy of the given molecules, LiH and BeH$_2$, respectively. For each molecule, the corresponding Hamiltonian is obtained as follows. First, the one- and two-body integrals in molecular-orbital basis are computed using the PySCF classical computational-chemistry software package, configured to use the STO-3G basis.  Next, these integrals are used to construct a qubit operator.  The qubit mapping chosen for this procedure is Jordan-Wigner. Following the process outlined by Setia \textit{et al.} in 2019, the number of qubits is tapered proportionally to the spatial symmetries present in the molecule under study~\cite{Setia2019}. Such qubit-tapering procedure is precision-preserving; it reduces the required qubits to simulate BeH$_2$ from 14 to 7, and LiH from 12 to 6, without performing any approximation. More details can be found in Appendix~\ref{appendix:hamiltonian-preparation}. Figure~\ref{fig:lih} compares the performance of EVQE and VQE/UCCSD at various interatomic distances. Figure~\ref{fig:beh2} compares the performance of EVQE and VQE/UCCSD at one specific interatomic distance. While beyond the scope of the comparison to VQE/UCCSD, for the sake of completeness, the appendix explores the performance of the VQE heuristic variational forms RyRz and Ry in the estimation of the ground-state energies of LiH and BeH$_2$.

\begin{table}[t]
\captionsetup{font=small}
    \centering
    {\renewcommand{\arraystretch}{1.3}
    \begin{tabular}{llll} 
        \toprule
        \vspace{1mm}
        \textbf{Noise Setting} & \textbf{Depth}$^*$ & \textbf{CX Count}$^*$  & \textbf{Error}$^*$      \\ 
        \hline
        0                      & 5.12$\pm$0.85    & 2.02$\pm$0.30       & 0.00357$\pm$0.00754  \\
        1                      & 2.36$\pm$1.91    & 0.66$\pm$0.96       & 0.05617$\pm$0.04143  \\
        2                      & 2.28$\pm$1.92    & 0.60$\pm$0.98       & 0.05887$\pm$0.04897 \\
        3                      & 1.90$\pm$1.61    & 0.52$\pm$1.09       & 0.06013$\pm$0.04796  \\
        4                      & 2.10$\pm$1.67    & 0.55$\pm$0.85       & 0.06067$\pm$0.05332  \\ 
        \hline
        \multicolumn{3}{l}{Optimal Separable State Average Error:}      & 0.06681              \\
        \bottomrule
        \multicolumn{4}{l}{$^*$Ranges reported represent one standard deviation from the mean.} \\
        \bottomrule
    \end{tabular}
    }
    \caption{\textbf{Randomized Testing: Variable-noise Experiment for 10 Random 2-qubit Hamiltonians.} EVQE is tested in noisy simulation using the device profile of the IBMQ Tokyo quantum computer in the estimation of the ground-state energy of 10 randomly generated Hamiltonians, with varying levels of noise present. As the noise settings decrease from 4 to 3 to 2 and finally to 1, the device's longitudinal-coherence and transverse-coherence times are multiplied by factors 1, 5, 10, and 100 respectively. At noise setting 0, only shot noise is present. The circuit executions for all algorithms are conducted using 1300 shots. The reported statistics come from the average of 5 trials for each of the 10 random Hamiltonians. EVQE is configured with a population size of 50, an optimization count of 150, an $\alpha$ of $10^{-3}$, a $\beta$ of $10^{-3}$, and the SPSA optimizer. The ``Optimal Separable State Energy Average Error'' corresponds to the average error associated with a circuit containing only U3 gates (no entanglement), whose parameters are optimized in ideal state-vector simulation utilizing an unbounded number of optimization iterations. Thus, it represents the maximum recoverable energy without entanglement.}
    \label{fig:noisyrand2qubit}
\end{table}

    In the two optimization problems presented, the task is to find the optimal solution of a small, randomly generated, NP-Hard problem instance. In these applications, the problems are encoded as Ising Hamiltonians using procedures similar to those outlined by Lucas in 2014~\cite{Lucas2014}.
    
\begin{figure*}[t]
\captionsetup{font=small}
    \centering
    \subfloat{\includegraphics[width=0.48\linewidth]{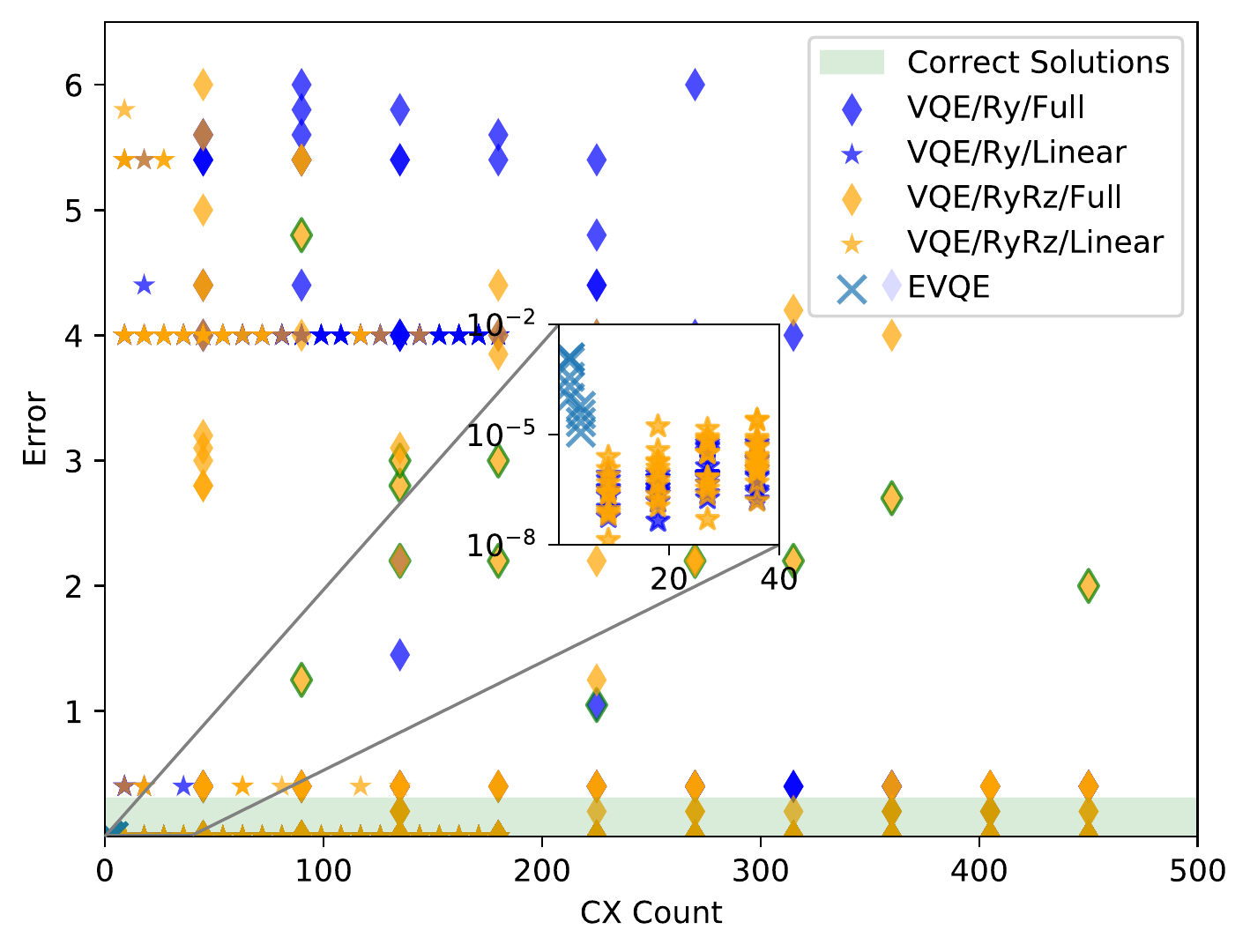}}\hfill
    \subfloat{\includegraphics[width=0.48\linewidth]{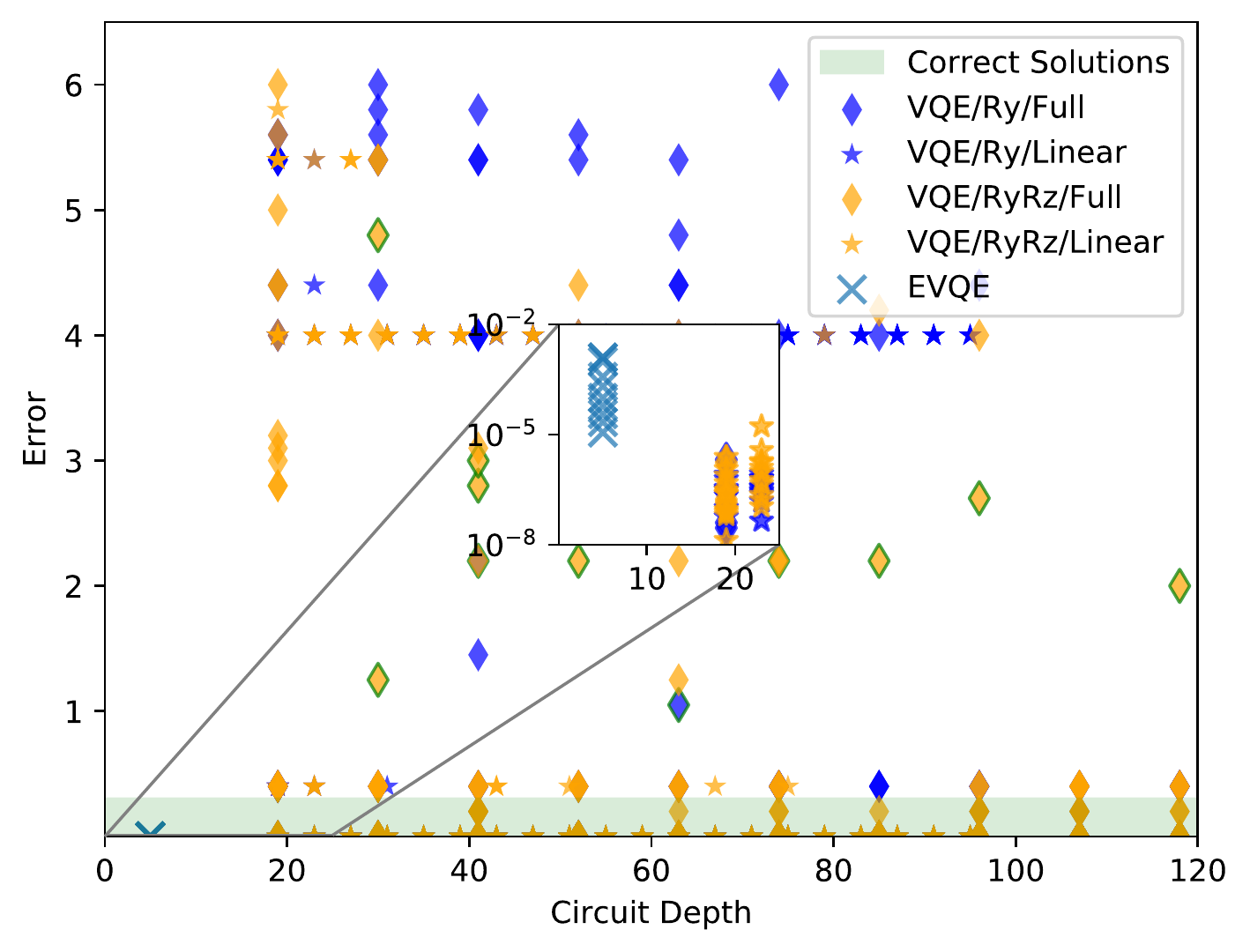}}
    \caption{\textbf{Optimization Application: State-vector Implementation for 10-qubit Max-Cut.} EVQE is compared against VQE in the calculation of the maximum cut (Max-Cut) of a random graph with 10 nodes. Each algorithm is tested 10 times, with all results shown above. Solutions that fall within the green shaded region correspond to optimal solutions of the random Max-Cut instance. Solutions with green outlines outside of the shaded region also correspond to optimal solutions. EVQE is configured with a population size of 250, an optimization count of 140, an $\alpha$ of $10^{-2}$, a $\beta$ of $1.25\times 10^{-3}$ and the COBYLA optimizer. For all variational forms, VQE demonstrates the best performance when using the SLSQP optimizer. EVQE is configured to yield solutions with approximately $10^{-2}$ error, while VQE's optimizer is configured with a tolerance of $10^{-7}$ to prevent premature convergence.}
     \label{fig:maxcut}
\end{figure*}

\subsection{Hardware Simulation}

The purpose of the hardware-simulation experiments is to understand the noise-resistance characteristics of EVQE, and to compare EVQE and VQE under realistic noisy conditions. As these experiments are designed to understand the algorithms' intrinsic noise-resistance properties, we do not use error mitigation techniques, such as those presented by Temme \textit{et al.} in 2017 and Kandala \textit{et al.} in 2019~\cite{Temme2017, Kandala2019}. All experiments are conducted using the Qiskit Aer QASM Simulator, configured with the noise profile and qubit connectivity of two IBMQ devices: 20-qubit Tokyo and 5-qubit Vigo. 

The purpose of the experiments shown in Table~\ref{fig:noisyrand2qubit} is to understand how the behavior of EVQE changes when various levels of noise are present. The noise characteristics are varied by scaling the hardware's T1 and T2 times, representing the longitudinal-coherence and transverse-coherence times, respectively, in simulation. As the coherence times increase, the effective noise decreases. The experiments are conducted on 10 randomly generated 2-qubit Hamiltonians, which are created using the \texttt{random\_hermitian} method in Qiskit Aqua \cite{Qiskit}. These experiments are conducted on 2 qubits as a matter of practicality as noisy simulation is computationally expensive. The noise profile and qubit connectivity correspond to the IBMQ Tokyo device. The characteristics of this device, such as the single-qubit- and CX-gate fidelity, are discussed by Cross \textit{et al.} in 2019~\cite{Cross2019}. As IBMQ Tokyo has 20 qubits, and the experiments only require 2 qubits, the algorithm is pinned to the 2 connected qubits with the lowest CX-gate error. The noise characteristics of this quantum computer require the use of a noise-resistant optimization subroutine, so the \textit{Simultaneous Perturbation Stochastic Approximation} (SPSA) optimizer is used for these experiments.

The purpose of the experiment shown in Figure~\ref{fig:noisylih} is to demonstrate EVQE's noise resistance relative to VQE in a practical application: the estimation of the ground-state energy of LiH. To allow the use of Vigo---one of IBM's latest 5-qubit quantum devices, with substantially improved single-qubit- and CX-gate fidelity---the simulation of LiH is configured by removing unoccupied spin orbitals, thereby reducing the number of required qubits from the 6 used in state-vector simulation down to just 4. Excluding unoccupied spin orbitals from the computation of the ground-state energy of LiH leads to negligible imprecision, which does not prevent chemical accuracy from being attained. In fact, this procedure has been used in various experiments \cite{Kandala2017, Kandala2019}. Furthermore, by running on IBMQ Vigo, EVQE is able to use the noise-sensitive optimizer COBYLA as its optimization subroutine. The device's noise profile also enables the use of COBYLA with VQE. However, in this experiment, VQE demonstrates the best performance when using SPSA. Finally, the configurations for EVQE and VQE are such that the total number of circuit executions performed by both algorithms are comparable. By controlling the total number of circuit evaluations, each algorithm's obtained error is a function of its noise resistance rather than the performance of the optimizer. This is in contrast to the state-vector simulations, where ideal parallelism is assumed, and thus EVQE is configured so that the total number of sequential circuit evaluations is comparable to the total number of circuit evaluations performed by VQE. Additionally, the decoherence constraints of hardware simulation prevent VQE/UCCSD from obtaining good results. Specifically, in this experiment, VQE/UCCSD obtains an error of $0.602\pm 0.026$ Hartree. Thus, for the purpose of examining the impacts of noise, EVQE is compared against VQE using the best heuristic variational form found. 

\begin{figure*}[]
\captionsetup{font=small}
    \centering
    \vspace{1cm}
    \subfloat{\includegraphics[width=0.48\linewidth]{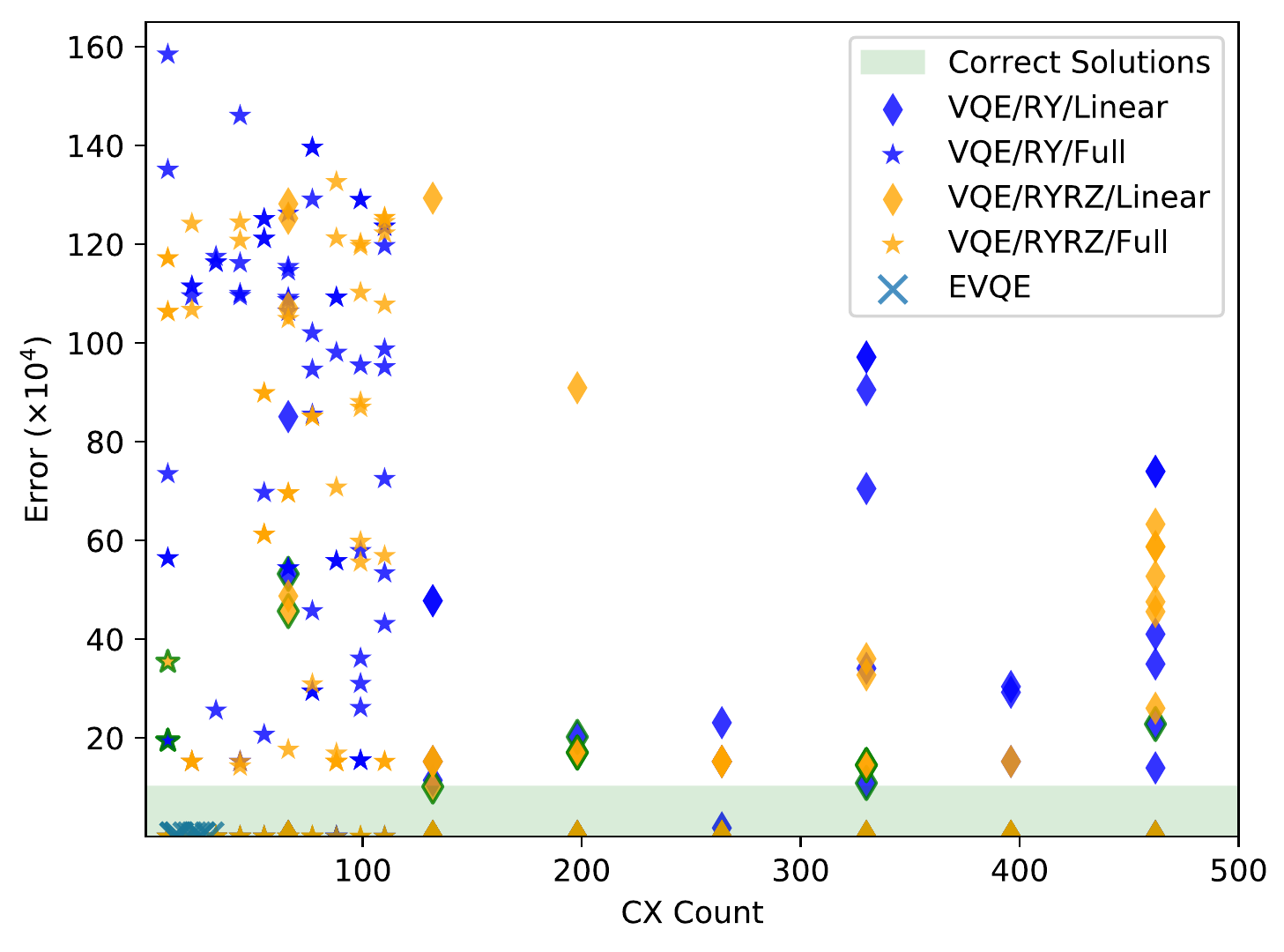}}\hfill
    \subfloat{\includegraphics[width=0.48\linewidth]{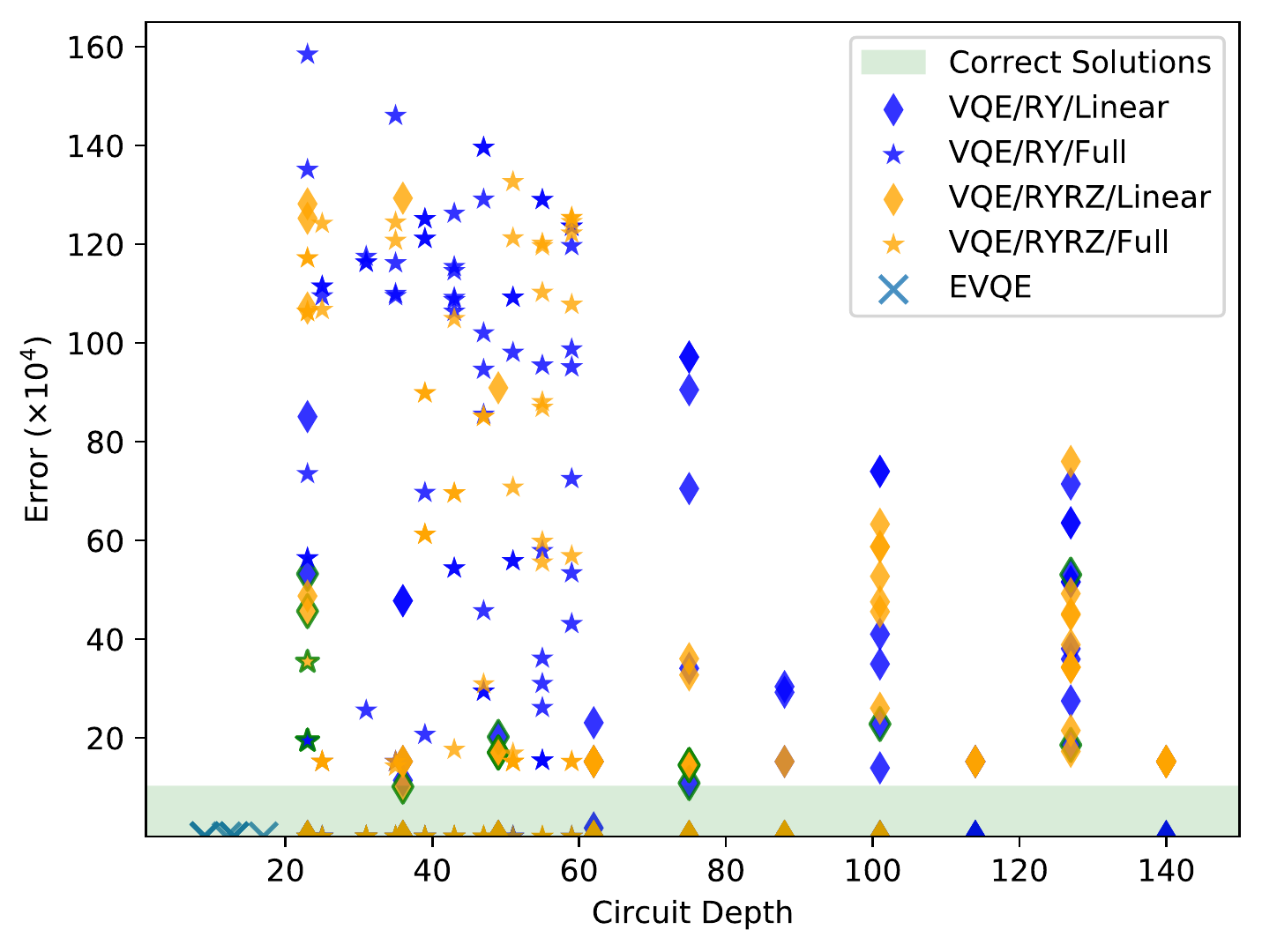}}
    \caption{\textbf{Logistics Application: State-vector Implementation for 12-qubit Vehicle Routing.} EVQE is compared against VQE in the calculation of an optimal set of routes for a fleet of 32 vehicles to deliver goods to 4 distinct destinations. Each algorithm is tested 10 times, with all results shown above. Solutions that fall within the green shaded region correspond to optimal solutions of the random vehicle-routing instance. Solutions with green outlines outside of the shaded region also correspond to optimal solutions. EVQE is configured with a population size of 300, an optimization count of 300, an $\alpha$ of $10^{-3}$, a $\beta$ of $10^{-2}$ and the COBYLA optimizer. For all variational forms, VQE demonstrates the best performance when using the SLSQP optimizer. EVQE is configured to yield solutions with an error of approximately $10^{-2}$, while VQE's optimizer is configured with a tolerance of $10^{-7}$ to prevent premature convergence.}
    \label{fig:vehiclerouting}
\end{figure*}

\subsection{Execution on Quantum Hardware}
The purpose of this experiment is to demonstrate the efficacy of EVQE when executed on real quantum hardware. We observe that for variational optimization to succeed, contiguous and relatively uninterrupted access to quantum hardware is likely necessary, as the natural drift in qubit calibration over time can change the mapping performed by a given circuit parameterization. Consequently, the experiment presented here---a random instance of Max-Cut with 5 vertices---is performed after having obtained exclusive access on 5-qubit IBMQ Vigo. 

For this experiment, EVQE is configured with a population size of 8, an optimization iteration limit of 50, a generation cap of 2, an $\alpha$ of $5\times 10^{3}$, a $\beta$ of $10^{-6}$, and the COBYLA optimizer. Such a conservative configuration is necessary to enable the algorithm to run during hardware exclusive allocation time. VQE is configured with the layouts RyRz/linear and Ry/linear as they demonstrate the best results in noisy simulation. Both VQE configurations use the COBYLA optimizer, set with a maximum of 500 optimization iterations. VQE is tested with depth settings ranging from 0 to 2. Both EVQE and VQE are executed with 1300 shots.

\section{Discussion}
When comparing the performance of variational algorithms, four important metrics  are the following: error, circuit depth, CX-gate count, and total number of circuit evaluations. These metrics vary in significance depending on the type of experiment being conducted.

In state-vector simulation, circuit depth and CX counts are important indicators of performance on actual hardware. CX gates typically have errors that are one order of magnitude larger than single-qubit gates. Therefore, their quantity is representative of the cumulative gate-fidelity error that will be incurred. This metric has been used as a gauge for circuit cost in various papers~\cite{Hu2019, Nam2019}. Additionally, the depth of a circuit is related to the required execution time.  Thus, it is an indirect measure of the decoherence error that would be incurred when the algorithm is executed on real quantum hardware. In real-hardware execution, as well as in hardware simulation, the impacts of CX-gate count and circuit depth are implicitly taken into account in the execution of a circuit. As a circuit's depth and CX-gate count increase, gate-fidelity and decoherance errors accumulate. Consequently, circuit executions yield results with less fidelity, increasing the error of the solution. Therefore, in noisy execution or simulation, total error is the primary metric of interest. In both state-vector simulation and real-hardware execution, the total number of circuit evaluations is an important consideration. However, as discussed in Section \ref{section:experimental_evaluation}, the total number of circuit evaluations is not directly comparable between EVQE and VQE. 

\subsection{State-vector Simulation}

It is worth emphasizing that for the results shown in Figures \ref{fig:lih}, \ref{fig:beh2}, \ref{fig:maxcut} and \ref{fig:vehiclerouting}, when near-ideal parallelism is assumed, EVQE's gradual circuit growth and parameter optimization result in the number of \textit{sequential circuit evaluations} (circuit evaluations that cannot be performed in parallel) being comparable to the total number of circuit evaluations performed by VQE when using variational forms that obtain similar levels of error. Practically, this means that when both algorithms are executed on a 64-core processor, EVQE generally has similar, if not faster, running times than VQE in the aforementioned experiments.

\begin{figure*}[]
\captionsetup{font=small}
    \centering
    \subfloat{\includegraphics[width=0.6\linewidth]{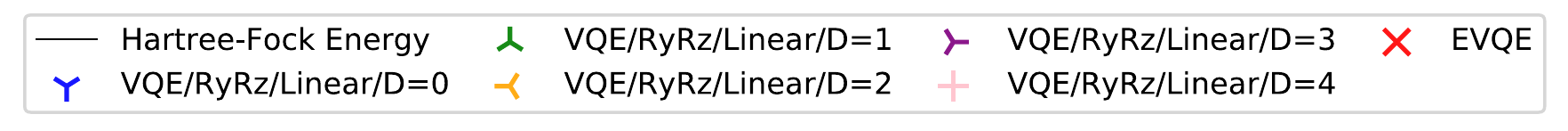}}\vspace{-2mm}\\%
    \subfloat{\includegraphics[width=0.4\linewidth]{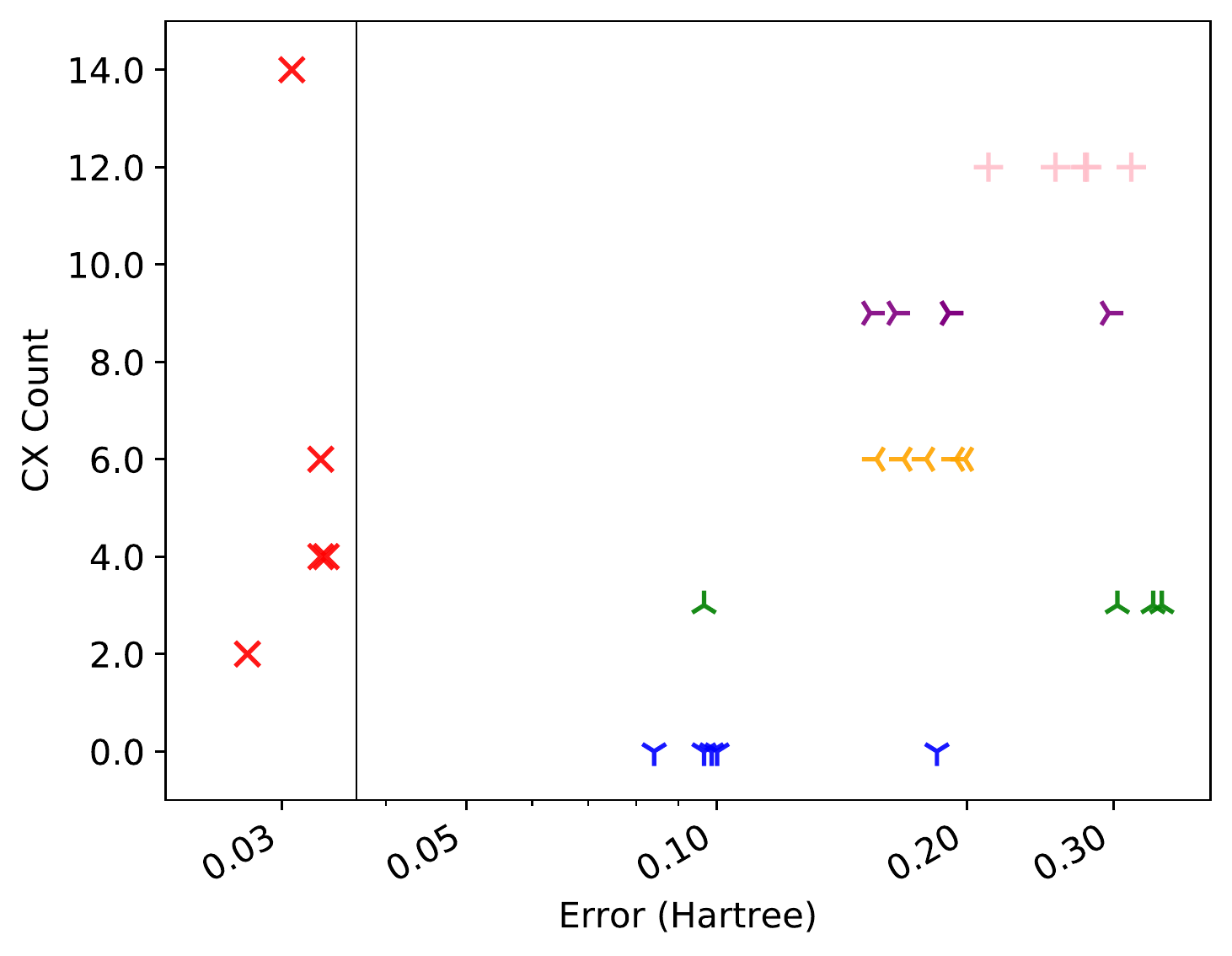}}\hspace*{3mm}%
    \subfloat{\includegraphics[width=0.4\linewidth]{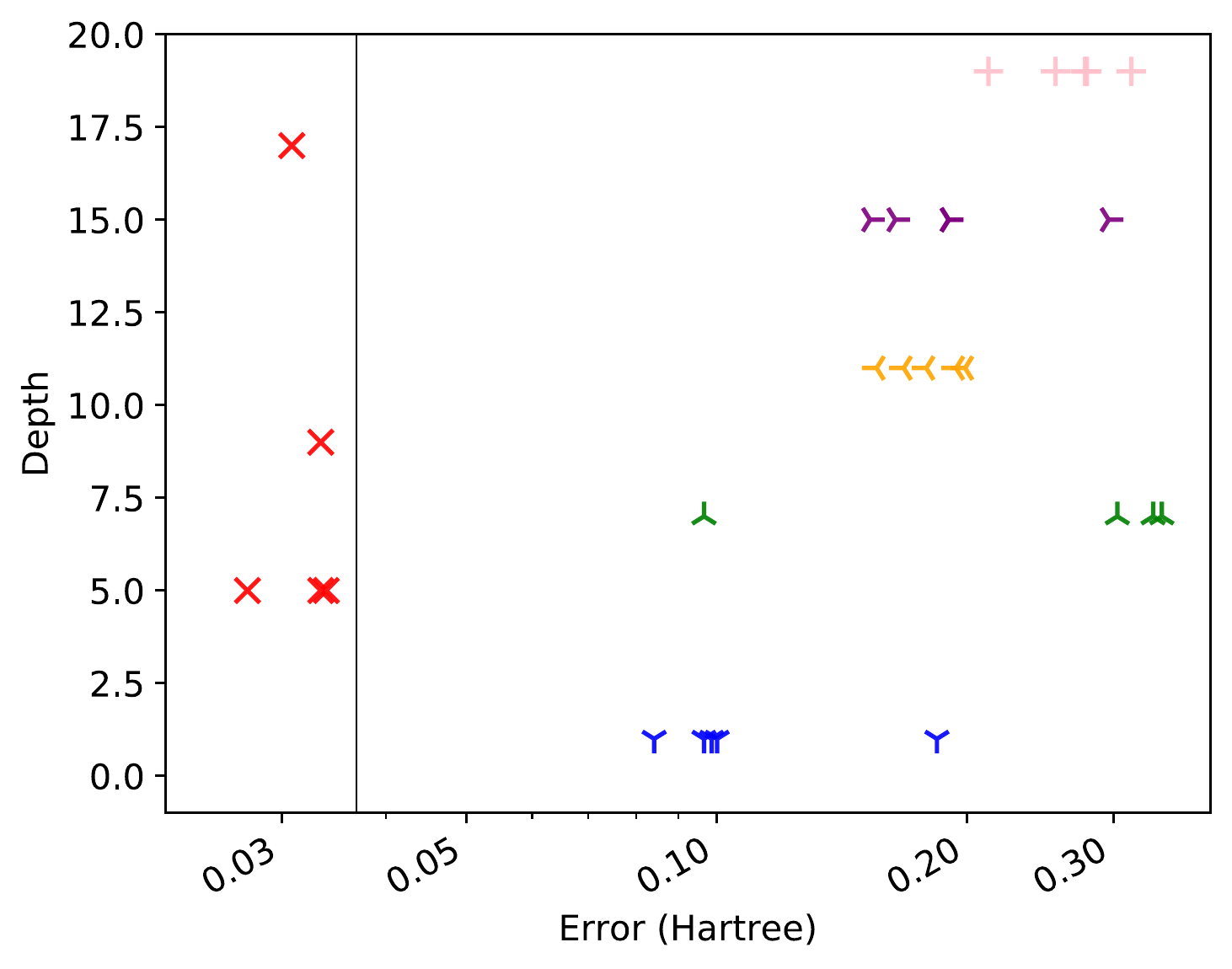}}
    \caption{\textbf{Chemistry Application: Hardware Simulation of IBMQ Vigo for 4-qubit LiH.} EVQE is compared against VQE in the calculation of the ground-state energy of LiH with an interatomic distance of $2.2\angstrom$, and unoccupied orbitals removed. In this experiment, the Qiskit Aer QASM simulator is configured with the noise profile and device connectivity of the IBMQ Vigo quantum computer, while EVQE is configured with a population size of 8, an optimization iteration count of 55, the COBYLA optimizer, an $\alpha$ of $0.005$ and a $\beta$ of $10^{-5}$. VQE is tested with the following variational-form/entanglement configurations: RyRz/linear with depths 0-4, RyRz/full with depths 0-4, Ry/linear with depths 0-4, Ry/full with depths 0-4, and UCCSD with depth 1. VQE/RyRz/linear gives the best results for VQE, and so is shown in the figures above. VQE performs best when configured with the SPSA optimizer. An optimization-iteration limit corresponding to 2040 circuit evaluations is set. The circuit executions for all algorithms are conducted using 1800 shots. All algorithms, apart from VQE/UCCSD, use the vacuum state as the starting state. VQE/UCCSD uses the Hartree-Fock state as its initial state.}
    \label{fig:noisylih}
\end{figure*}

The chemistry state-vector simulations demonstrate significant resource reductions for EVQE compared to VQE/UCCSD. Figure \ref{fig:lih} shows the circuit depths and CX counts required for EVQE and VQE/UCCSD to obtain chemical accuracy in the estimation of the ground-state energy of LiH at 10 different interatomic distances. The same configuration is used for each algorithm at all interatomic distances. On average, EVQE ranges from having $5.0\times$ to $15.2\times$ shallower circuits than VQE/UCCSD, while using $3.5\times$ to $5.0\times$ fewer CX gates.  It is important to notice that \textit{this result in EVQE is obtained without using any domain-specific variational form}. As expected from an algorithm that dynamically grows its circuits, EVQE develops deeper circuits with more CX gates at interatomic distances for which the estimation of the ground state is more challenging, as opposed to VQE/UCCSD, which has constant resource usage at all interatomic distances. Similarly, Figure \ref{fig:beh2} shows the circuit depths and CX counts required for EVQE and VQE/UCCSD to obtain chemical accuracy in the estimation of the ground state energy of BeH$_2$ at an interatomic distance of $1.3\angstrom$. Relative to the LiH experiment, EVQE is configured with a larger population size as well as a larger optimization-subroutine iteration limit. Consequently, an even larger reduction in resources is observed relative to VQE/UCCSD, with EVQE obtaining an $18.6\times$ shallower circuit, using $12.0\times$ fewer CX gates. The results obtained in both of the chemistry state-vector experiments are consistent with the hypothesized properties of EVQE. 

Figures \ref{fig:maxcut} and \ref{fig:vehiclerouting} compare EVQE and VQE in the calculation of optimal solutions to two NP-Hard optimization problems. In both experiments, EVQE's circuits are always shallower than the shallowest possible solutions for VQE (when two body operators are used). Additionally, EVQE's CX counts are comparable to the minimum required by any of the VQE configurations. In these experiments, EVQE always obtains the optimal solution while a significant number of VQE's solutions, up to 90\% for some configurations, obtain sub-optimal results.

EVQE's consistency is likely a result of three of its properties. First, its optimization scheme only optimizes one circuit layer at a time, according to a random order. Therefore, EVQE's optimization subroutine only has $O(n)$ parameters to concurrently optimize. This is in contrast to VQE, which simultaneously optimizes all of the parameters in its circuits, and so for a variational circuit with depth $d$, has $O(dn)$ parameters to concurrently optimize. EVQE's second beneficial property is its population of circuits. If some circuits in the population are more susceptible to premature convergence than others, over a number of generations the population will tend towards the circuits that can be optimized more effectively. Additionally, if the probability of prematurely converging is modelled as a random variable, since the population contains multiple species of similar circuits, the probability of all of the circuits in a species converging prematurely is exponentially small in terms of the size of the specie. EVQE's third advantageous property is also enabled by speciation. VQE's premature convergence can occur when the algorithm arrives at a parameterization corresponding to a local, but not global, optimum in the search space. In contrast, EVQE's speciation ensures that the circuits in its population are distributed among various peaks in the search space, reducing the likelihood that all the circuits in the population converge to local, non-global, optimum. It is worth noting that VQE was tested with a variety of optimizers, including SLSQP, COBYLA, and SPSA, and premature convergence was still observed. Moreover, in Figure \ref{fig:maxcut}, EVQE generally obtains higher error than the VQE results falling in the correct solution region. This is by design of the experiment.  In fact, EVQE was configured by setting the depth and $\land_1(\textup{U3})$ penalties to obtain solutions with error of magnitude $10^{-2}$. In contrast, to prevent premature convergence, all VQE configurations were configured to converge with a tolerance of $10^{-7}$, well explaining the results. If a solution with greater accuracy is desired for EVQE, it may be obtained by reducing the magnitude of the depth and $\land_1(\textup{U3})$ penalties. 

\subsection{Hardware Simulation}

Table~\ref{fig:noisyrand2qubit} examines the noise-resistance properties of EVQE. As the noise setting increases, the effective noise experienced by the algorithm also increases. At around noise setting 2, corresponding to transverse and longitudinal coherence times $5\times$ greater than in the actual device profile, the algorithm's error and resource usage begin to plateau. As the error increases, the algorithm tends towards using fewer CX gates, which is as expected given that CX gates have lower fidelity and longer execution times than U3 gates. Additionally, as the average number of CX gates decreases, the algorithm's average error approaches the average error of the optimal separable state energy. This energy represents the minimum energy of the Hamiltonian without taking into account correlation energy, and so this behavior is consistent with the fact that, as the number of CX gates in a circuit tends to zero, the circuit's energy evaluation approaches this limit. As single-qubit gates have relatively fast execution times and relatively high fidelity, it follows that circuits with only a single layer of U3 gates would very closely approximate the ideal separable energy. EVQE experiences a sharp increase in error when increasing the noise setting from 0 to 1. At noise setting 0, only shot noise is present. However, at noise setting 1, both decoherence \textit{and} gate-fidelity error are present, the latter of which is constant at noise settings 1, 2, and 3. Had the gate fidelity error been varied as well, a more gradual increase in error would likely have been observed. This experiment demonstrates that the algorithm will automatically tend towards the circuits that obtain the minimum energy evaluation possible given the constraints imposed upon the forms of plausible circuits by the noise characteristics of the device being used. Possible extensions of this experiment could repeat it with higher qubit counts, or by decreasing the T1 and T2 times substantially past their actual values, thereby emulating conditions that are noisier than real hardware.

Figure \ref{fig:noisylih} compares EVQE and VQE in hardware simulation in the practical application of computing the ground-state energy of LiH. EVQE obtains errors that range from being $3.4\times$ to $11.6\times$ smaller than that obtained by VQE using the best variational form. Moreover, no VQE configuration obtains an energy smaller than the Hartree-Fock energy, while all trials with EVQE do. \textit{In the presence of noise, only EVQE obtains results accounting for some of the correlation energy of the molecule}. VQE's performance tends to get worse as the depth setting increases. As the transpiled depth and CX count increase, the effective noise experienced by VQE also increases, resulting in noisier energy evaluations. 

It is somewhat surprising that VQE's results do not improve from depth setting 0 to depth setting 1, as depth setting 1 introduces three CX gates, granting VQE the ability to recover correlation energy. It is possible that VQE/RyRz/linear requires multiple layers of its variational circuit pattern to be repeated in order to capture the entanglement present in the molecule. In contrast, EVQE does not repeat a certain circuit pattern to increase its depth. Rather, it automatically evolves and adapts its circuits. Thus, EVQE can find circuits with sufficient depth, containing CX gates capturing the necessary entanglement, which minimize its energy evaluations despite the presence of noise.

Originally, VQE was executed using the COBYLA optimizer. However, COBYLA consistently converged to average errors ranging from 0.11 to 0.26 Hartree after approximately 250 circuit evaluations, and so the SPSA optimizer was used instead as it obtained better results, as SPSA fully executes the user-specified number of optimization iterations. In an attempt to prevent this problem, COBYLA was configured with various termination tolerances, up to $10^{-25}$ in magnitude. While this termination threshold more than tripled the average number of circuit evaluations performed, no meaningful improvements to the calculated error were observed. Other optimizers such as SLSQP and Nelder-Mead were also tested. All configurations of SLSQP failed to obtain errors below $0.4$ Hartree for VQE/RyRz with a depth setting of 0, and so it was not tested with other circuit depths. Nelder-Mead obtained comparable results to COBYLA, while conducting substantially more circuit evaluations. While it is hard to rule out that the difference in performance between EVQE and VQE is not entirely the product of the optimizers, we can certainly conclude that it is very difficult to find an optimizer configuration for VQE that yields results similar to those obtained by EVQE.

EVQE's bottom four data points in the figure (noting that two data points overlap) behave as expected: they describe relatively shallow circuits, with a reasonable number of CX gates, and yield good energy evaluations. Surprisingly, however, EVQE finds a circuit with a depth of 16 and with 14 CX gates, which also obtains a low error of 0.0308 Hartree. Given the results from all the VQE trials, we would not have expected a circuit with this number of CX gates and this circuit depth to obtain such a result. To confirm that random noise was not responsible for such a low energy evaluation, and that the output circuit indeed corresponds to a good solution, a new experiment is conducted where the circuit is tested in noiseless state-vector simulation; in this case, the error is 0.0391 Hartree. 

\subsection{Execution on Real Quantum Hardware}

Finally, the results of the hardware execution are presented. Given the requirement of exclusive hardware access, each algorithm configuration is only tested once, and so these results are illustrative rather than demonstrative. After the second generation, EVQE's fittest genome had an error of $0.059055$ and a circuit depth of 5, and used 2 CX gates. It obtained the optimal solution in $97.4\%$ of the 1300 shots taken. Out of the VQE configurations tested, only RyRz/linear with depth 0 obtained the optimal solution as the state with highest probability, obtaining it in $81.4\%$ of shots. The error in the energy evaluation associated with this solution is $0.077424$. Given the size of the problem instance, it follows that a VQE variational form with no CX gates, and with a transpiled depth of 1, could obtain the optimal solution. However, the poor performance of the other variational form configurations remains unclear. 

\section{Conclusion}
In this paper, we presented the Evolutionary Variational Quantum Eigensolver (EVQE), a novel variational algorithm that uses evolutionary programming techniques to minimize the expectation value of a given Hamiltonian by dynamically generating and optimizing an ansatz. EVQE has important new characteristics compared to other variational quantum algorithms, such as VQE:
\begin{enumerate}
    \item EVQE performs accurate energy evaluations with hardware-efficient ansatzes.  The circuits generated by EVQE in molecular simulations are up to $18.6\times$ shallower and use up to $12\times$ fewer CX gates than those generated by VQE coupled with domain-specific variational forms.
    \item Due to its evolutionary nature, EVQE operates in a domain-agnostic fashion, making it equally applicable to problems in diverse fields, such as chemistry, optimization, finance and artificial intelligence, thereby eliminating the need for domain-specific variational forms (whose construction requires specialized knowledge) or heuristic variational forms (which severely constrain the exploration of the Hilbert space and consequently are often inefficient and imprecise).
    \item EVQE is quantum-hardware adaptive; it automatically favors the construction of circuits that are more resilient to the noise characteristics and connectivity constraints of the specific quantum computer on which the algorithm is executed, obtaining results in noisy simulation with at least $3.6\times$ less error than VQE using \textit{any} tested ansatz configuration.
\end{enumerate}
EVQE demonstrates the potential of such an adaptive approach, motivating further study into the topic. Possible extensions of this work include comprehensive analyses of the noise-resistance properties of the algorithm, studies confirming that the seeding of the initial population coupled with identity-initialized growth circumvent the barren-plateau problem, and the exploration of custom optimization subroutines tailored to the dynamic growth patterns specific to EVQE. Additionally, exploring novel adaptive algorithms that use fewer total circuit evaluations may prove important while exclusive access to quantum hardware remains a rare resource.  

\section{Author Contributions}
AGR conceived the project. AGR, SH and MP designed and tested the algorithm. All the authors contributed both to the writing of the article and the development of the software used to test and demonstrate the algorithm. RC and SW contributed to the integration of EVQE into Qiskit Aqua.

\section{Acknowledgements}
We would like to thank Sergey Bravyi, Jay Gambetta, Abhinav Kandala, Julia Rice, Rahul Sarkar, Kanav Setia and Andrew Wack for the various discussions and suggestions.

\printbibliography

\vspace{0.5cm}

\section{Appendix}

\subsection{Identity-initialized Growth}
To illustrate the importance of identity initialized growth, we present the results from an experiment finding the ground state energy of a randomly generated 4 qubit Hermitian matrix. In \textit{EVQE-Standard}, EVQE is configured as described in the paper: when new gates are added to the circuit, be it U3 or $\land_1(U3)$ gates, their parameters are set such that they perform the identity transformation. In \textit{EVQE-CX} the algorithm is unchanged, except instead of adding identity initialized $\land_1(U3)$ gates, CX gates are added instead. All other parameters are held constant. However, this experiment does not control for the difference in parameter count, given that each $\land_1(U3)$ gate in EVQE-Standard has three parameters (as opposed to the non-parameterized CX gate used by EVQE-CX). 

By using the CX gate, the energy evaluations of the circuits produced by EVQE-CX can change non-smoothly, and according to our claim, should be of detriment to the algorithm's rate of convergence. The rate of convergence is analysed by comparing the most fit genome in each population after each generation. The algorithm which obtains lower error with fewer generations is considered to converge more quickly. 

\begin{figure}[t]
\captionsetup{font=small}
    \centering
    \includegraphics[width=0.9\linewidth]{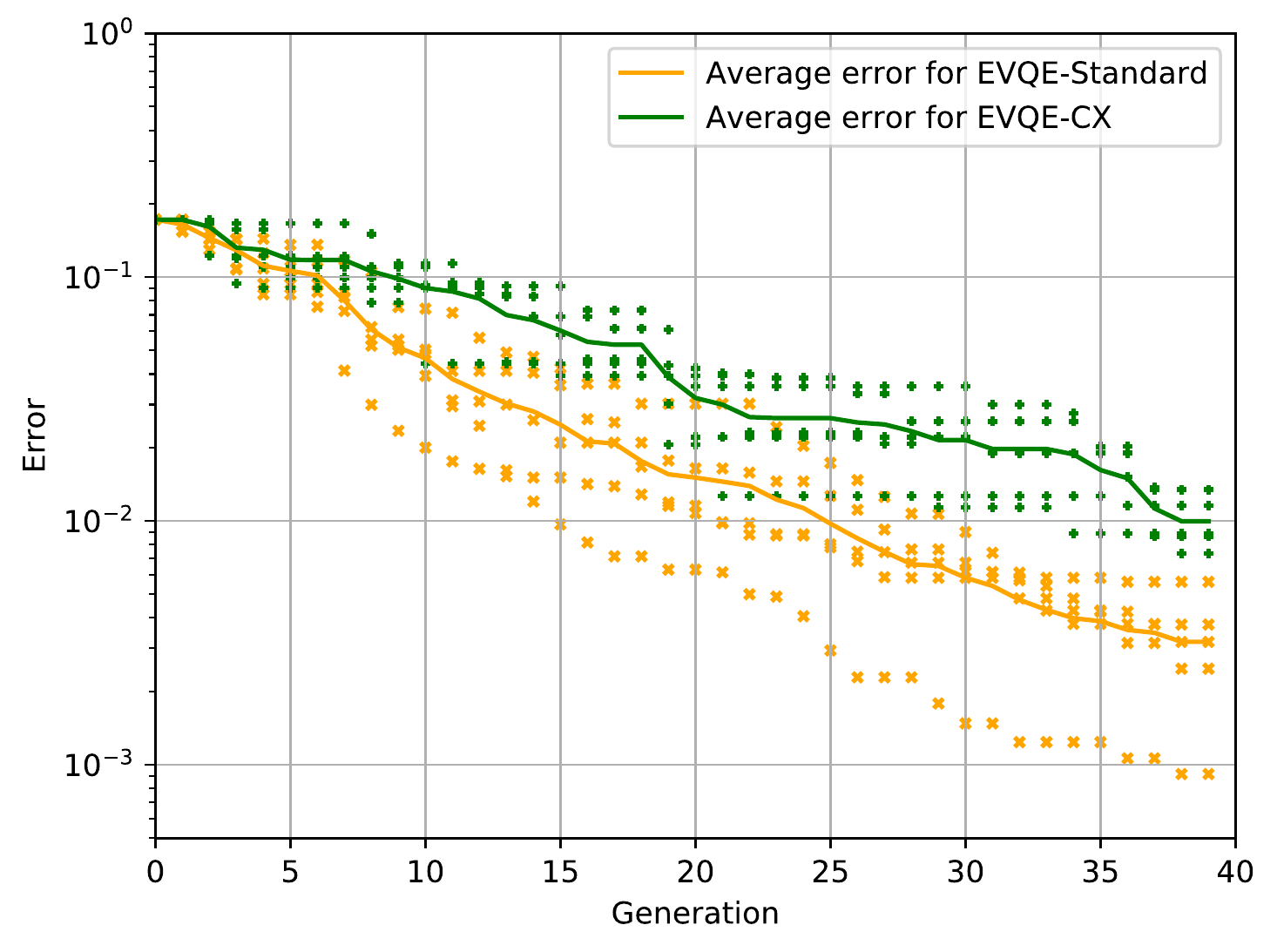}
    \caption{\textbf{Performance Analysis: State-Vector Evaluation of Identity Initialized Circuit Growth.} A random 4-qubit Hermitian matrix is generated using the \texttt{random\_hermitian} method in Qiskit Aqua \cite{Qiskit}. Its ground state energy is then calculated utilizing EVQE-Standard, which is the algorithm described in the paper, and EVQE-CX, which does not use identity initialized growth and grows circuits by adding CX gates rather than identity initialized $\land_1(U3)$ gates. Apart from the method of circuit growth, both algorithms have identical configurations: a population size of $25$, an optimization iteration count of $100$, the COBYLA optimizer, an $\alpha$ of $5\times 10^{-6}$, and to ensure a fair comparison, a $\beta$ of $0$.} 
    \label{fig:cx_vs_cu3}
\end{figure}

The experiment shown in Figure \ref{fig:cx_vs_cu3} supports our claim, demonstrating that identity initialized circuit growth results in error that is on average half an order of magnitude less than that obtained by non-gradient preserving circuit growth. Furthermore, we predict that this advantage becomes greater as qubit count increases. While it is difficult to see in the figure, individual trials for EVQE-CX can get stuck at given error rates for a number of consecutive generations (in some cases, exceeding 10 generations), before they suddenly improve by chance and the process repeats. By contrast, EVQE-Standard tends to constantly decrease its error without getting stuck at any specific error level for a significant number of generations. Identity initialized growth enables the algorithm to consistently improve its fitness evaluation with the addition of every circuit layer. However, as already mentioned, since EVQE-Standard utilizes parameterized $\land_1(U3)$ gates, and EVQE-CX utilizes non-parameterized CX gates, EVQE-Standard's genomes tend to have more parameters than the genomes in EVQE-CX. Consequently, when all other variables are fixed, EVQE-Standard tends to perform more circuit evaluations per generation than EVQE-CX. 

\subsection{State-vector Simulation of BeH$_2$ and LiH Using Heuristic Ansatzes}

In Figures \ref{fig:lih} and \ref{fig:beh2} of Section \ref{section:experimental_evaluation}, EVQE is compared against a variational form, UCCSD, explicitly designed for molecular simulation. By using UCCSD, there is some intuition ensuring that results obtained are close approximations of the true ground state energies. Thus, VQE/UCCSD is a good standard against which to compare variational algorithms, and has been used as such a benchmark in other papers such as ADAPT-VQE \cite{Grimsley2019}. However, just as with the NP-Hard problems presented in Section \ref{section:experimental_evaluation}, it is also possible to perform molecular simulations utilizing fixed heuristic variational forms, with no intuition or guarantees that they are even able to capture the required transformations. Given the problems with fixed variational forms outlined in Section \ref{section:introduction}, we do not expect such approaches to effectively or efficiently scale to higher qubit counts, where the volume of states captured by a fixed variational form is exponentially small in relation to the total space. However, for completeness, here we include the results from the best heuristic variational forms tested, and compare them against the results presented in Figures \ref{fig:lih} and \ref{fig:beh2}.

\begin{figure}[]
\captionsetup{font=small}
    \centering
    \includegraphics[width=0.84\linewidth]{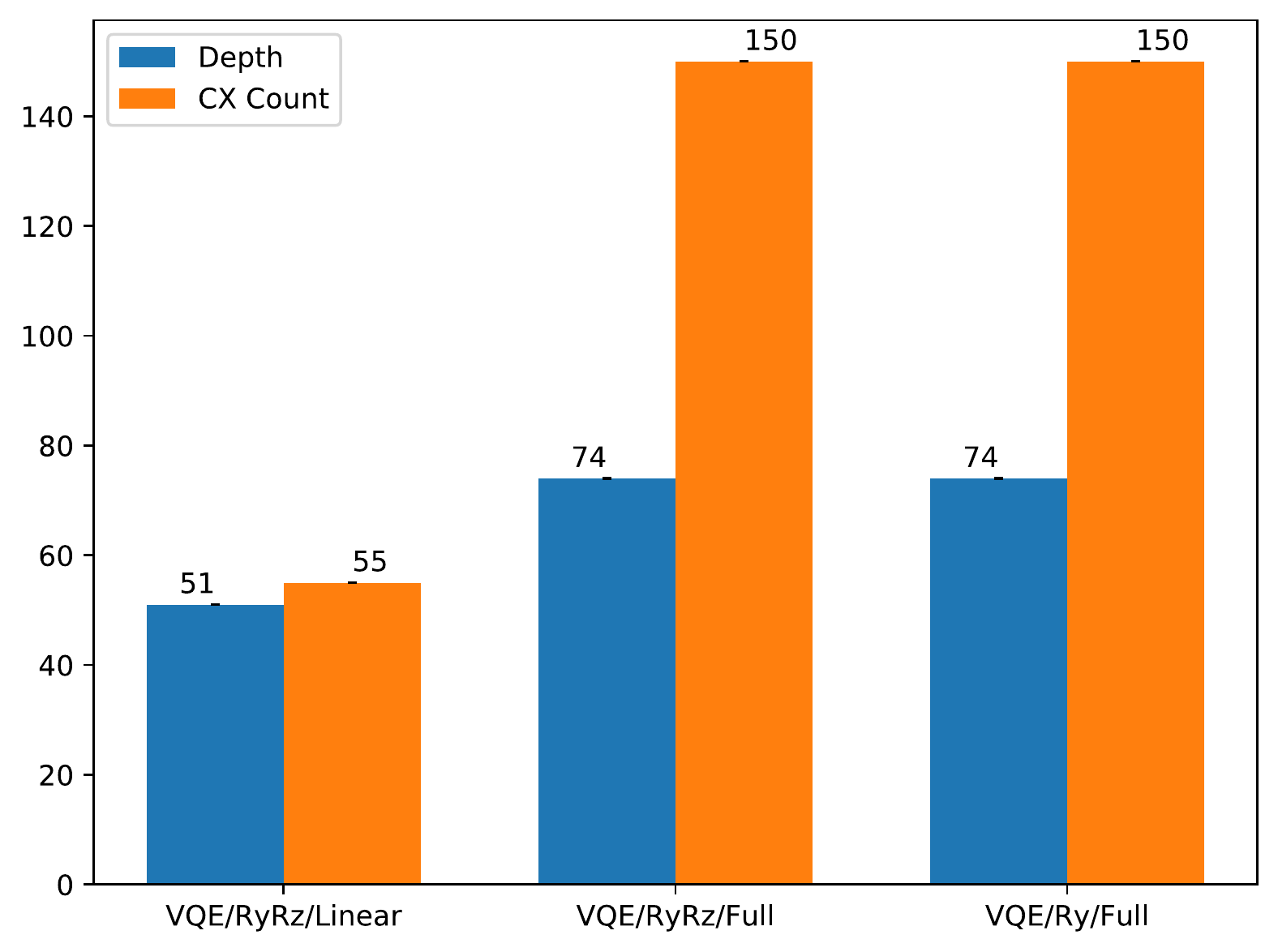}
    \caption{\textbf{Chemistry application: State-Vector Implementation for 6-qubit LiH Utilizing Heuristic Ansatz.} This figure represents the same experiment, with the same configurations, as that shown in Figure \ref{fig:lih}. The depths and CX count shown correspond to the minimum for each variational form configuration which, on the average of five trials, obtain chemical accuracy in the estimation of the ground state energy of LiH with the 10 interatomic distances shown in Figure \ref{fig:lih}. Each of these VQE heuristic variational forms demonstrate the best performance when using the SLSQP optimizer with an unbounded number of optimization iterations, terminating only upon convergence.}
    \label{fig:lih_appendix}
\end{figure}

In Figure~\ref{fig:lih_appendix}, the results from VQE with the RyRz variational form and linear entanglement (VQE/RyRz/linear), VQE with the Ry variational form and full entanglement (VQE/Ry/full), and VQE with RyRz and full entanglement (VQE/RyRz/full) using the minimum depth and CX counts that obtain chemical accuracy at all tested interatomic distances (on the average of 5 trials) are shown. (Here, we should clarify that the entanglement policy is said to be \textit{linear} if each qubit is only entangled with its neighbor, and \textit{full} if each qubit is entangled with every other qubit.) To determine these values, each algorithm was tested at each depth setting, in increasing order, until chemical accuracy was obtained on average. Moreover, each of these configurations were tested with a number of optimizers (such as COBYLA and SLSQP) to determine which yielded the best performance. The variational form configuration Ry/linear was also tested, but none of its tested configurations were able to obtain chemical accuracy at all interatomic distances. This configuration process was manually intensive and required a large number of circuit evaluations. Even so, EVQE obtained chemical accuracy with shallower circuits and fewer CX gates than all heuristic configurations of VQE, apart from VQE/RyRz/linear at the interatomic distances of $3.0\angstrom$ and $3.4\angstrom$ where the heuristic variational form had marginally shallower circuits with marginally fewer CX gates. It is likely that increasing EVQE's population size or the iteration count would result in EVQE outperforming all conceivable configurations of the aforementioned heuristic ansatzes.

\begin{figure}[]
\captionsetup{font=small}
    \centering
    \includegraphics[width=0.84\linewidth]{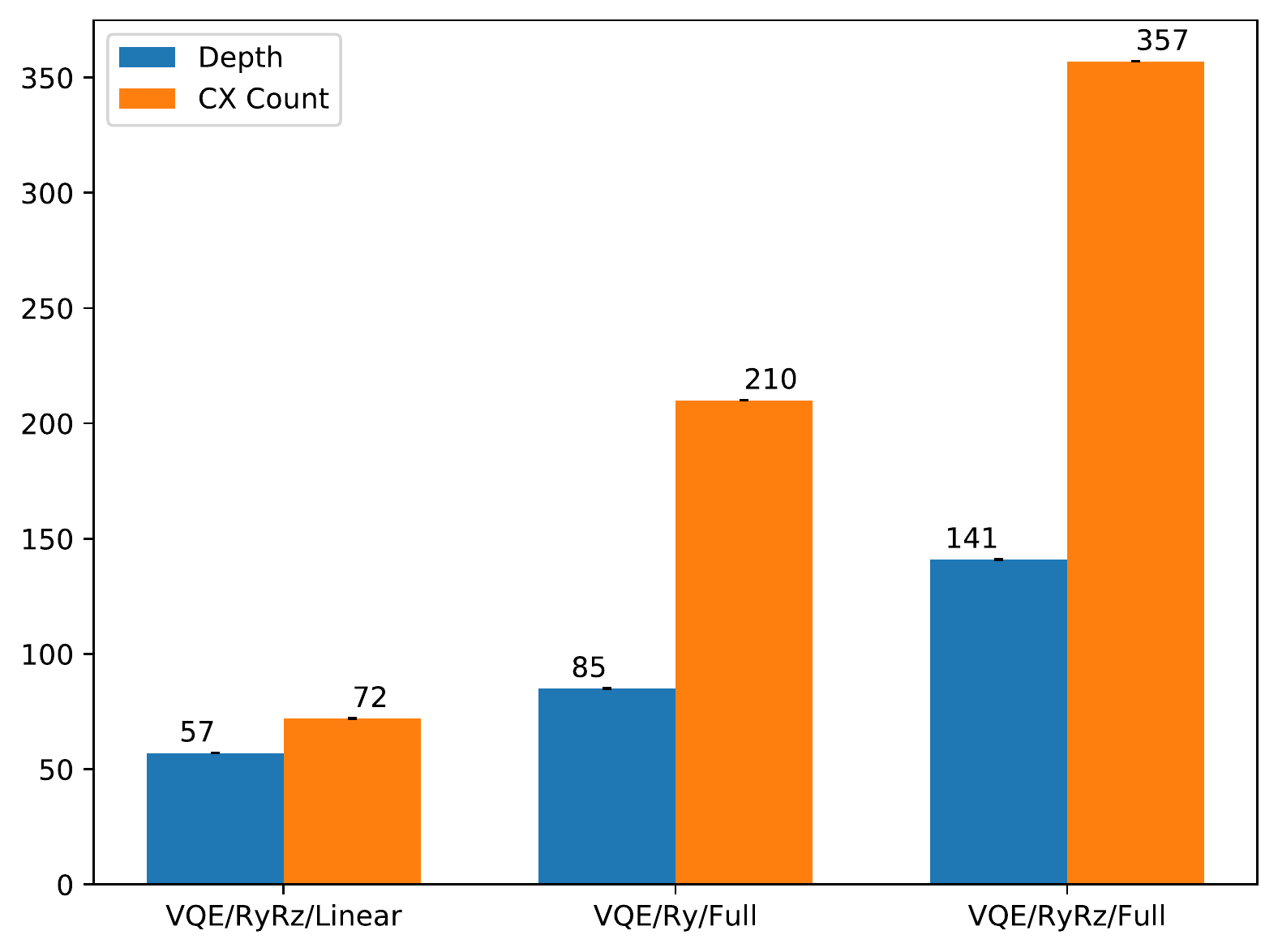}
    \caption{\textbf{Chemistry Application: State-Vector Implementation for 7-qubit BeH$_2$ Utilizing Heuristic Ansatz.} This figure represents the same experiment, with the same configurations, as that shown in Figure \ref{fig:beh2}. Each of these VQE heuristic variational forms demonstrate the best performance when using the SLSQP optimizer with an unbounded number of optimization iterations, terminating only upon convergence. The depth setting used for each heuristic form was the minimum obtaining chemical accuracy over the average of 10 trials.}
    \label{fig:beh2_appendix}
\end{figure}

The results shown in Figure \ref{fig:beh2_appendix} were obtained by configuring the heuristic variational forms in much the same way as just discussed for LiH. Again, the Ry/linear variational form configuration was unable to obtain chemical accuracy. One significant difference for these experiments is that no heuristic configuration of VQE that obtained chemical accuracy on average used fewer resources than EVQE. That is, in the calculation of the ground-state energy of BeH$_2$, EVQE obtained chemical accuracy while using fewer resources than \textit{any} configuration of VQE that was tested. The relative favorable performance of EVQE in the BeH$_2$ experiments is likely a result of one of, or a combination of, two factors. First, the population size used in the experiment shown in Figure \ref{fig:beh2} was larger than that used in Figure \ref{fig:lih}. Second, as predicted, heuristic variational forms may struggle as the number of qubits increase.

\subsection{Pruning the Space of Circuit Forms}
\label{appendix:pruning-optimizations}

One additional technique that we utilize to optimize our proposed evolutionary strategy is the careful pruning of the search space, as guided by the following set of rules when constructing genomes during EVQE's evolutionary process. Specifically, these rules are applied when creating a new random gene to add to a given genome.

\begin{itemize}
    \item No qubit will have two $\textup{U3}$ gates applied to it in succession. The reason should be quite obvious as two consecutive $\textup{U3}$s may be contracted and represented as a single $\textup{U3}$. Having them as two separate $\textup{U3}$s increases the challenge to the classical optimizer as six ``free'' parameters (for the two $\textup{U3}$s) would have to be tuned to find the optimum as opposed to three (for the single $\textup{U3}$).
    
    \item Similarly, we also do not allow any pairs of qubits to have two consecutive $\textup{CU3}$s applied to them, unless the control and target are reversed between the two gates. Again this follows from observing that two consecutive $\textup{CU3}$ gates with the same target and controls may be contracted into a single $\textup{CU3}$ gate with a different parameterization. 
    
    \item One additional rule has to do with how the identity gate, $\mathbb{I}_2$, is used in genome constructions---they are only added if neither a $\textup{U3}$ nor a $\textup{CU3}$ gate can be added, in order to reduce the dimension of the parameter space. This implies that no $\mathbb{I}_2$ gate can immediately follow a $\textup{CU3}$ gate, or directly precede a $\textup{U3}$ gate.
\end{itemize}

Our prototype implementation of EVQE was constructed with all of the above pruning rules, helping to reduce the complexity of the search space, thereby making the evolutionary process more efficient.

\subsection{Qubit Hamiltonian Preparation}
\label{appendix:hamiltonian-preparation}
Here we describe how the Hamiltonians are generated for the experiments, including precision preserving 6-qubit LiH, precision preserving 7-qubit BeH$_2$, and approximated 4-qubit LiH. 
First, the atoms are aligned in the x-axis for both molecules, so all major interactions happen in x-axis only. This represents an unexploited optimization, as the algorithm could have been implemented with controlled $R_z$ gates instead of $\land_1(U3)$ gates. Moreover, for BeH$_2$, inter-atomic distance refers to the distance between the Be and H atoms. The molecular Hamiltonians are computed with STO-3G basis set, widely used in quantum chemistry, with the PySCF chemistry driver. For those three molecular Hamiltonians, we also freeze the $1s$ orbitals of Li and Be since they do not strongly interact with other orbitals. Furthermore, we either explore the molecular symmetries present or remove unoccupied orbitals to further reduce the number of qubits. Subsequently, we apply the Jordan-Wigner (JW) transformation to convert molecular Hamiltonians to qubit Hamiltonians. Note that, the JW transformation maps one fermionic mode onto one qubit.

\textbf{Precision-preserving 6-qubit LiH.} Under the above setting, LiH has 10 fermionic modes, and thus requires 10 qubits. We subsequently explore the symmetries in the molecule to further taper qubits based on the works of Bravyi \textit{et al.} and Setia \textit{et al.}~\cite{bravyi2017tapering, Setia2019}. For LiH, this approach can find four symmetries, resulting in a 6-qubit LiH. Bravyi's symmetry finding package is available in Qiskit Aqua~\cite{Qiskit}.

\textbf{Precision-preserving 7-qubit BeH$_2$.} Under the above setting, BeH$_2$ has 12 fermionic modes, and thus requires 12 qubits. Following the same procedure detailed for precision-preserving 6-qubit LiH, five symmetries are found, resulting in a 7-qubit BeH$_2$. 

\textbf{Approximated 4-qubit LiH} Under the above setting, LiH has 10 fermionic modes. Furthermore, we remove the weakly interacting orbitals ($2p_y$ and $2p_z$) since the atoms are aligned in the x-axis. Thus, 4 ferminonic modes are removed, resulting in a molecular Hamiltonian with 6 modes. Subsequently, we further explore the symmetries in the molecule, now finding 2 additional symmetries. Finally, only 4 qubits are required to approximate LiH.

\end{document}